\documentclass[pre,aps,final,twocolumn,showpacs,eqsecnum]{revtex4-1}
\newlength\figwidth\figwidth=0.48\textwidth

\usepackage{color,graphicx}
\usepackage{amsmath,amssymb,bm}
\usepackage{colordvi}

\def\bboxI{\mathbf{I}}
\def\bboxT{\mathbf{\Theta}}

\def\bfr{\mathbf{r}}
\def\bfp{\mathbf{p}}
\def\Re{\mathop{\rm Re}}

\def\Tr{\mathop{\rm Tr}}
\def\erf{\mathop{\rm erf}}
\def\rmd{\mathrm{d}}

\def\ramaSM{\vadjust{\vbox to 0pt{\vss \hbox to \hsize
{\hskip\hsize \quad $\Leftarrow$\quad {\it SM}\hss}\vskip3.5pt}}}
\def\ramaKA{\vadjust{\vbox to 0pt{\vss \hbox to \hsize
{\hskip\hsize \quad $\Leftarrow$\quad {\it KA}\hss}\vskip3.5pt}}}

\begin{document}

\title{SEMICLASSICAL TRACE FORMULA FOR THE TWO-DIMENSIONAL
 RADIAL POWER-LAW POTENTIALS}

\author{A.\ G.\ Magner\footnote{magner@kinr.kiev.ua}}
\affiliation{
 Institute for Nuclear Research, 03680 Kiev, Ukraine}
\author{A.\ A.\ Vlasenko}
\affiliation{
 Institute for Nuclear Research, 03680 Kiev, Ukraine}
\affiliation{
 Institute of Physics and Technology, NTUU ``KPI", 03056 Kyiv, Ukraine}
\author{K.\ Arita}
\affiliation{
 Department of Physics, Nagoya Institute of Technology,
 Nagoya~466-8555, Japan}

\begin{abstract}
The trace formula for the density of single-particle levels in the
two-dimensional radial power-law potentials,
which nicely approximate up to a constant shift 
the radial dependence of the Woods-Saxon
potential and its quantum spectra in a bound region, was derived by the
improved stationary phase method.  The specific analytical results are
obtained for the powers $\alpha=4$ and 6.  The enhancement 
of periodic-orbit contribution to the level density
near the bifurcations are found to be significant for the description
of the fine shell structure.
The semiclassical trace formulas for the shell corrections to the level
density and the energy of many-fermion systems reproduce
the quantum results with good accuracy through all the bifurcation
(symmetry breaking) catastrophe points, where the standard
stationary-phase method breaks down.  Various limits (including the
harmonic oscillator and the spherical billiard) are obtained from the
same analytical trace formula.
\end{abstract}


\date{\today}

\maketitle

\section{Introduction}

According to the shell-correction method (SCM) \cite{str2,fuhi}, the
oscillating part of the total energy of a finite fermion system, the
so-called shell-correction energy $\delta U$, is associated with an
inhomogeneity of the single-particle  
energy level distributions near the Fermi
surface.  Depending on the level density at the Fermi energy -- and
thus the shell-correction energy $\delta U$ -- being a maximum or a
minimum, the many-fermion system is particularly unstable or stable,
respectively.  Therefore, the stability of this
system varies strongly with particle numbers and
parameters of the mean-field potential and external force.

A semiclassical periodic orbit theory (POT) of shell
effects\cite{gutz,strm,bt,crlit} was used for a deeper
understanding, based on classical pictures, of the origin of nuclear
shell structure and its relation to a possible chaotic nature of the
dynamics of nucleons.  This theory provides us with a nice tool for
answering, sometimes even analytically, the fundamental questions
concerning the exotic physical phenomena in
many-fermion systems; for instance, the origin of
the double-humped fission barrier
and, in particular, of the creation of the isomer minimum
in the potential energy surface
\cite{strd,brreisie,spheroid,migdal,book}.  
Some applications of the POT to nuclear deformation energies
were presented and discussed for the infinitely deep potential
wells with sharp edges 
in relation to the bifurcations of periodic orbits (POs)
with the pronounced shell effects.

In the way to more realistic semiclassical calculations, it is
important to account for a diffuseness of the nuclear edge.  It is
known that the central part of the realistic effective mean-filed
potential for nuclei or metallic clusters are described by the
Woods-Saxon (WS) potential $V_{\rm WS}(r)$ \cite{WS}.  The idea of
Refs.~\cite{aritapap,arita2012} is that the WS potential is nicely
approximated (up to a constant shift) by much a simpler power-law
potential which is proportional to a power of the radial coordinate
$r^\alpha$.  The approximate equality
\begin{equation}
V_{\rm WS}(r) \approx V_{\rm WS}(0) + W_0 r^\alpha
\label{ramod}
\end{equation}
holds up to around the Fermi energy with a suitable choice of the
parameters $W_0$ and $\alpha$.  In the case of the spatial dimension
${\cal D}= 2$, one can use Eq.~(\ref{ramod}) for a realistic
potential of electrons in a circular quantum dot
\cite{migdal,book,reimann96}.  We shall derive first the generic trace
formula for this radial power-law (RPL) potential in the case of two
dimensions, and then discuss its well known limits to the harmonic
oscillator and cavity (billiard) potentials \cite{book}. 
The main focus will be aimed to the non-linear dynamics depending on
the power parameter $\alpha$ to show the symmetry-breaking 
(bifurcation) phenomena. They lead to the remarkable enhancement 
of PO amplitudes of the level density and energy shell corrections 
which was found within the improved stationary phase
approximation (improved SPM, or simply ISPM) 
\cite{ellipse,spheroid,maf,migdal}. The ISPM means more exact evaluation 
of the trace formula
integrals with the finite limits over a
classically 
accessible phase-space volume and with higher-order (if necessary) expansions 
of the action phase
of the exponent and pre-exponent factors up to the first non-zero terms
with respect to the standard SPM (SSPM) \cite{gutz,strm,bt,crlit}. 
In this way, one may remove the SSPM 
discontinuities and divergences.

The manuscript is organized as follows.  In Sec.~\ref{sec2} the
classical dynamics is specified for the RPL potentials.
The trace formulas
for the RPL potentials in two dimensions are derived in Sec.~\ref{sec3}.
Section~\ref{sec4} is devoted to the comparison of the semiclassical
calculations for the oscillating level density and  shell-correction
energy with quantum results.  The paper is summarized in
Sec.~\ref{sec5}.  Some details of our POT calculations, in particular
full analytical derivations at the powers $\alpha=4$ (see also
Ref.~\cite{trap}) and $6$ for all POs and those 
at arbitrary $\alpha$
for the diameter and circle orbits, are given in the 
Appendixes~\ref{appA}--\ref{appE}.

\section{Classical dynamics and bifurcations}
\label{sec2}

The radial power-law (RPL) potential model is described by the
Hamiltonian
\begin{equation}\label{ramodra}
H=\frac{p^2}{2m}+E_0\left(\frac{r}{R_0}\right)^\alpha,
\end{equation}
where $m$ is the mass of the particle; $R_0$ and $E_0$ are introduced
as constants having the dimension of length and energy, respectively,
and are related with $W_0$ in Eq.~(\ref{ramod}) by
$W_0=E_0/R_0^\alpha$.  (In practice, we fix $E_0$ and adjust the
WS potential by varying $R_0$ and $\alpha$.)
This Hamiltonian includes the limits of the harmonic
oscillator $(\alpha=2)$ and the cavity $(\alpha\to\infty)$; realistic
nuclear potentials with steep but smooth surfaces correspond to values
in the range $2<\alpha<\infty$.  The advantage of this potential is
that it is a homogeneous function of the coordinates, so that the
classical equations of motion are invariant under the scale
transformations:
\begin{gather}
\bfr\to s^{1/\alpha}\bfr, \quad
\bfp\to s^{1/2}\bfp, \quad
t \to s^{1/\alpha-1/2}t \nonumber \\
\text{with $E\to sE$}\;. \label{scaling}
\end{gather}
Therefore, one only has to solve the classical dynamics once at a
fixed energy, e.g., $E=E_0$ ($s=1$); the results for all other
energies $E$ are then simply given by the scale transformations
(\ref{scaling}) with $s=E/E_0$ by definition in the last equation of
Eq.~(\ref{scaling}).  This highly simplifies the POT
analysis \cite{aritapap,bridge}.
Note that the definition
(\ref{ramodra}) can also be generalized to include deformations (see,
e.g., Ref.~\cite{bridge,arita2012}).
 
As we consider the spherical RPL Hamiltonian (\ref{ramodra}), it can
be written explicitly in the two-dimensional (2D) spherical canonical
phase-space variables $\{r,\varphi;p_r,p_\varphi\}$,
where $\varphi$ is the azimuthal angle (a cyclic variable),
$p_\varphi=L$ is the angular momentum, and the radial momentum
$p_r$ is given by
\begin{align}
p_r(r,L) &= \sqrt{p^2(r)-\frac{L^2}{r^2}}\;, \nonumber \\
&\quad p(r)=\sqrt{2m\left[E-E_0\left(\frac{r}{R_0}\right)^\alpha\right]}\;.
\label{pr}
\end{align}
The classical trajectory (CT) $r(t)$ can be easily found by
integrating the radial equation of motion $\dot{r}=p_r/m$ with
Eq.~(\ref{pr}).  Transforming the spherical canonical variables into
the action-angle ones, for the actions $I_r,I_\varphi$ one has
\begin{gather}
I_r=
\frac{1}{\pi}\,\int_{r_{\rm min}}^{r_{\rm max}}p_r\,\rmd r\;
\equiv I_r(E,L),\label{actionvar}\\
I_\varphi=\frac{1}{2\pi}\,\int_0^{2\pi} p_\varphi \,\rmd\varphi \equiv L\;,
\end{gather}
where $r_{\rm min}$ and $r_{\rm max}$ 
are the turning points which are the two real (positive) solutions of the 
equation $p_r^2(r,L)=0$.

\begin{figure*} 
\begin{center}
\includegraphics[width=1.4\figwidth]{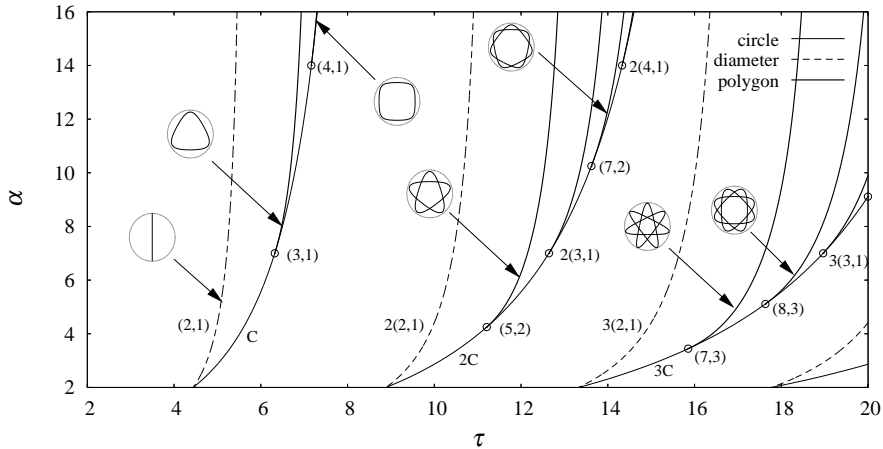}
\end{center}
\caption{
Scaled periods $\tau_{{}_{\!\rm PO}}$ of some short
POs as functions of the power parameter $\alpha$
in dimensionless units $m=R_0=E_0=1$ (Appendix~\ref{appB}).
Thin solid curves are the circle orbits $MC$,
dashed curves are the diameters $M(2,1)$, and thick solid curves are
the polygon-like orbits $M(n_r,n_\varphi)$ $(n_r>2n_\varphi)$; 
their bifurcations from the $MC$ are indicated by open circles.}
\label{fig1}
\end{figure*}

The definition (\ref{ramodra}) can be used in arbitrary spatial
dimensions, as long as $r$ is the corresponding radial variable. In
practice, we are interested only in the 2D and 3D cases.  The
spherical 3D and the circular 2D potential models have common PO sets, 
see Fig.~\ref{fig1}.  For $\alpha>2$, POs with the
highest degeneracy [$\mathcal{K}=1$ (3) in the 2D (3D) cases] are
specified by three integers and labeled as $M(n_r,n_\varphi)$,
where $n_r$ and $n_\varphi$ are mutually 
commensurable
numbers of oscillations in the radial direction,
and of rotations around the
origin, each for a primitive orbit, respectively;
and $M$ is the repetition number.
For the isotropic harmonic oscillator ($\alpha=2$), all the classical
orbits are periodic ones with (degenerate) ellipse shapes.  By
slightly varying $\alpha$ away from 2, the specific diameter and
circle orbits appear separately, and they remain as the shortest POs
with the corresponding degeneracies $\mathcal{K}=1$ and $0$.  With
increasing $\alpha$, the circle orbit and its repetitions cause
successive bifurcations generating various new periodic orbits
$\{n_r,n_\varphi\}$, $ n_r > 2n_\varphi$.  Fig.~\ref{fig1} shows some
of the shortest POs $M(n_r,n_\varphi)$.  The shortest PO is the
diameter which has the degeneracy $\mathcal{K}=1$ in the 2D problem at
$\alpha>2$.  Other polygon-like orbits have $\mathcal{K}=1$ at $\alpha
> \alpha_{\rm bif}$, where $\alpha_{\rm bif}$ is a bifurcation value
(see its specific expression below).  The circle orbit having maximum
angular momentum is isolated ($\mathcal{K}=0$) for the 2D system
(except for the bifurcation points).

For the frequencies of the radial and angular motion of particle,
one finds
\begin{equation}
\omega_r=\frac{\partial H}{\partial I_r}
=\left(\frac{\partial I_r}{\partial E}\right)_L^{-1},\quad
\omega_\varphi=\frac{\partial H}{\partial L}=
-\frac{\left(\partial I_r/\partial L\right)_E}{
 \left(\partial I_r/\partial E\right)_L}\;,
\label{freq}
\end{equation}
where 
$I_r=I_r(E,L)$ [Eq.~(\ref{actionvar})] is identical to the energy
surface $H(I_r,L)=E$.  Thus, the PO condition is written as
\begin{equation}
f(L) \equiv \frac{\omega_\varphi}{\omega_r}=\frac{n_\varphi}{n_r}\;,
\label{percond}
\end{equation}
where
\begin{equation}
f(L) =-\left(\frac{\partial I_r(E,L)}{\partial L}\right)_E=
\frac{L}{\pi} \,\int_{r_{\rm min}}^{r_{\rm max}}\frac{\rmd r}
{r^2p_r(r,L)}\;.
\label{fldef}
\end{equation}
The energy surface $I_r=I_r(E,L)$
is simply considered as a function of
only one variable $L$ [Eq.~(\ref{actionvar})].
The solutions to
the PO equation [see Eq. (\ref{percond})], 
$L^*=L^*(n_r,n_\varphi)$, for the
given co-primitive integers $n_\varphi$ and $n_r$ define
the one-parametric families $\mathcal{K}=1$ of orbits $M(n_r,n_\varphi)$
because $L$ is the single-valued integral of motion, which is only
one (besides the energy $E$) in the 2D case\cite{strm,strd}.
The azimuthal angle $\varphi$ can be taken, for instance, as a parameter
of the orbit of such a family.

According to the limit $f(L) \rightarrow 1/2$ at $L \rightarrow 0$,
one has the diameter orbits $M(2,1)$ as the specific one-parametric
($\mathcal{K}=1$) families related to the solution $L=0$ of
Eq.~(\ref{percond}).  The other specific solutions are the isolated
($\mathcal{K}=0$) circle orbits $MC$ by which we represent the $M$-th
repetition of the primitive circle orbit $C$.  The radius $r_{{}_{\!C}}$
of the circle orbit is determined by the system of equations
$r_{\rm min}=r_{\rm max}\equiv r_{{}_{\!C}}$, or equivalently by
equations (\ref{eqrc}) (see Appendix~\ref{appA}).
Thus the angular momentum of the circle orbit is given by
$L_C=r_{{}_{\!C}}p(r_{{}_{\!C}})$.  As seen obviously from the 
condition of
the real radial momentum $p_r$ [Eq.~(\ref{pr})], this $L_C$ is the
maximal value of the angular momentum $L$, i.e.,
$0 \leq |L| \leq L_C\;$.

As shown in Appendix~\ref{appA}, for the
stability factor $F_{MC}$ of the circle orbit $MC$ in the radial
direction, defined in Refs.\ \cite{gutz,book} through 
the trace of the 
PO stability matrix, $\Tr(\mathcal{M}_{MC})$,
one obtains
\begin{align}
F_{MC}&=2-\Tr(\mathcal{M}_{C})^M = 
4 \sin^2\left[\frac{\pi M \varOmega_C}{\omega_{{}_{\!C}}}\right] \nonumber \\
&=4 \sin^2\left[\pi M \sqrt{2+\alpha} \right]\;, 
\label{fgutzC}
\end{align}
where $\varOmega_C$ [Eq.\ (\ref{omc})] and 
$\omega_{{}_{\!C}}$ [Eq.\ (\ref{omtcra})] are
the radial and angular frequencies of the circle orbit.
This factor $F_{MC}$ is zero at the bifurcation points
$\alpha_{\rm bif}$ by the definition of the stability
matrix, $\Tr(\mathcal{M}_C)^M=2$, for the POs 
($\varOmega_C/\omega_{{}_{\!C}}\equiv \sqrt{2+\alpha}=n_r/n_{\varphi}$),
\begin{equation}
\alpha_{\rm bif}=\frac{n_r^2}{n_\varphi^2}-2\;.
\label{bifeqra}
\end{equation}
The PO family
$M(n_r,n_\varphi)$, which corresponds to the solutions $L^*<L_C$ of the
PO equation (\ref{percond}), exists for all $\alpha >\alpha_{\rm bif}$.
There is the specific bifurcation point $\alpha=2$ in the spherical
harmonic oscillator (HO) limit with the frequency
$\omega_\varphi=\sqrt{2E_0/(mR_0^2)}$, 
where one has the two-parametric
families at any $L$ within a continuum $0 \leq L\leq
E/\omega_\varphi$.  In the HO limit, the above specified circle and
diameter orbits belong to these families.
In the circular billiard limit $\alpha \rightarrow \infty$, the
isolated circle orbit $(\mathcal{K}=0)$ is degenerating into
the billiard boundary $r_{{}_{\!C}} \rightarrow R_0$, $L_C \rightarrow
\sqrt{2mE}\,R_0$ [see the limit $\alpha\to\infty$ in
Eq.~(\ref{rcLcd2Fcra}) for $r_{{}_{\!C}}$].

Another key quantity in the POT is the curvature $K$  
of the energy surface $I_r=I_r(E,L)$\, given by
\begin{equation}
K=
\frac{\partial^2 I_r(E,L)}{\partial L^2}= 
-\frac{\partial f(L)}{\partial L}\,,
\label{curv}
\end{equation}
where $f(L)$ is the ratio of frequencies [Eq.~(\ref{fldef})].  
As shown below, the curvature (\ref{curv}) and
Gutzwiller factor (\ref{fgutzC})
are the key quantities for calculations of the magnitude of the
PO contributions into the semiclassical level density.

\section{Trace formulas}
\label{sec3}

The level density $g(E)$ for the Hamiltonian 
$H(\bfr,\bfp)$ can be obtained by 
using the phase-space trace formula (in $\mathcal{D}$ dimensions)
 \cite{ellipse,spheroid,maf,siebshom}:
\begin{gather}
g_{\rm scl}(E) = \frac{1}{(2\pi\hbar)^{\mathcal{D}}}\Re\sum_{\rm CT}
\int \rmd \bfr^\prime \int \rmd \bfp^{\prime\prime}\: 
\delta \left(E - H(\bfr^{\prime\prime},\bfp^{\prime\prime})\right)
\nonumber\\ \times
\left|\mathcal{J}_{\rm CT}(\bfp^\prime_\perp,\bfp^{\prime\prime}_\perp)\right|^{1/2}\;
\exp\left(\frac{i}{\hbar} \varPhi_{\rm CT}
- i \frac{\pi}{2} \mu_{{}_{\!\rm CT}}\right)\;. \label{pstrace}
\end{gather}
The sum is taken over all discrete CT manifolds
for a particle moving between the initial 
$\bfr^\prime,\bfp^\prime$; and the final
$\bfr^{\prime\prime},\bfp^{\prime\prime}$ points with a given energy $E$.
Any CT can be uniquely specified by fixing, for instance, the initial
condition $\bfr^\prime$, and the final momentum $\bfp^{\prime\prime}$
for a given time $t_{{}_{\!\rm CT}}$ of the motion along the CT.  For the
action phase $\varPhi_{\rm CT}$ in exponent of (\ref{pstrace}), one has
\begin{align}
\varPhi_{\rm CT} &\equiv S_{\rm CT}(\bfp^\prime,\bfp^{\prime\prime},
t_{{}_{\!\rm CT}}) +
\left(\bfp^{\prime\prime}- \bfp^\prime\right) 
 \cdot \bfr^\prime \nonumber \\ &= S_{\rm CT}(\bfr^\prime,
\bfr^{\prime\prime},E) -
\bfp^{\prime\prime} \cdot \left(\bfr^{\prime\prime} -
\bfr^\prime\right)\,,  
\label{legendtrans}
\end{align}
where 
$S_{\rm CT}(\bfp^\prime,\bfp^{\prime\prime},t_{{}_{\!\rm CT}}) = 
-\int_{\bfp^\prime}^{\bfp^{\prime\prime}}
\rmd \bfp \cdot \bfr(\bfp)\;$ and 
$S_{\rm CT}(\bfr^\prime,\bfr^{\prime\prime},E) = 
\int_{\bfr^\prime}^{\bfr^{\prime\prime}}
\rmd \bfr \cdot \bfp(\bfr)\;$ 
are the actions in the
momentum and coordinate representations, respectively.
In Eq.~(\ref{pstrace}), 
$\mathcal{J}_{\rm CT}(\bfp^\prime_\perp,\bfp^{\prime\prime}_\perp)$ is the Jacobian for
the transformation of the initial momentum 
$\bfp^\prime_\perp$ to the final one $\bfp^{\prime\prime}_\perp$
in the direction perpendicular to CT.
$\mu_{{}_{\!\rm CT}}$ is the Maslov phase related to
the number of conjugate (turning and caustics) points along the CT
\cite{fedoryukpr,maslov}. 

One of the terms in Eq.\ (\ref{pstrace}) is 
related to the local short zero-action CT which is the well known 
Thomas-Fermi (TF) level density \cite{maf,migdal}.
For calculations of the other oscillating terms of the 
trace integral (\ref{pstrace}),  one may use
the  ISPM, expanding the action phase 
$\Phi_{\rm CT}$ and pre-exponent factor in both
$\bfp^{\prime\prime}$ and $\bfr^\prime$ variables
up to the first non-zero terms with the finite integration 
limits over the classically
accessible phase-space region \cite{maf,migdal}. 
The {\it stationary phase} conditions are equivalent to the
{\it periodic-orbit} equations, and therefore,
the oscillating level density can be presented as the sum over POs 
in a potential well
\cite{migdal,book}.

\subsection{One-parametric orbit families $(\mathcal{K}=1)$ }
\label{sec3A}
In order to obtain the contribution of the  one-parametric   
{\it families} of the maximal degeneracy $\mathcal{K}=1$
into the phase-space trace formula (\ref{pstrace}),
it is useful to transform 
the usual Cartesian phase-space variables $\{\bfp;\bfr\}$ to the
other canonical action-angle ones $\{\bboxI;\bboxT\}$, specified in
the spherical action-angle variables as
$\bboxT=\{\varTheta_r,\varTheta_\varphi \equiv \varphi\}$;
$\bboxI=\{I_r,I_\varphi\equiv L\}$.
The Hamiltonian $H$, action phase $\varPhi_{\rm CT}$,  and other related
quantities of the integrand in Eq.~(\ref{pstrace})
[e.g. $H=H(\bboxI)=H(I_r,I_\varphi) \equiv H(I_r,L)$]  
are independent of the angle variables $\bboxT$.
Therefore, one can easily perform the integration
over these angle variables $\bboxT$, which gives
the factor $(2\pi)^2$.
Then, taking the integral
over $I_r$ exactly by using the energy conserving $\delta$-function,
for the oscillating terms of the CT sum (\ref{pstrace}), one obtains
\begin{align}
&\delta g_{\rm scl}(E)
= \frac{1}{2\hbar^2}\Re\sum_{M,n_r,n_\varphi}\, \int
\rmd L\:\frac{1}{\omega_r} \nonumber \\ &\times\exp\left\{\frac{2\pi i}{\hbar}
M\,\left[n_r\,I_r(E,L) + n_\varphi L\right]-
\frac{i \pi}{ 2}\;\mu_{{}_{\!M,n_r,n_\varphi}}
\right\}\;. \label{poissonsum}
\end{align}
Here, the phase (\ref{legendtrans}) is expressed in terms of the
corresponding action-angle variables through the actions  
in the considered mixed representation,
\begin{equation}
\varPhi_{\rm CT}=2\pi M\left[n_r\,I_r(E,L) + n_\varphi \,L\right],
\label{phase3}
\end{equation}
$n_r$ and $n_\varphi$ are {\it positive} co-primitive
integers, $M$ is a nonzero integer, $\omega_r$ is the radial frequency 
in Eq.~(\ref{freq}). 
We also omit the upper indexes in $\bboxI$  (or $\{I_r,L\}$) variables
which represent initial (prime) and final (double primes) values of
Eq.~(\ref{pstrace}), taking explicitly into account that these
variables are
constants of motion for the spherical integrable
Hamiltonian.  The integration limits in Eq.~(\ref{poissonsum}) 
for $L$ are $-L_C \leq L \leq L_C$, where $L_C$ is
the maximum value corresponding to the circle orbit.
All quantities in the integrand are taken at the energy surface
$I_r=I_r(E,L)$ [Eq.~(\ref{actionvar})].
Thus, Eq.~(\ref{poissonsum}) is similar to 
the oscillating component of the 
semiclassical Poisson summation trace formula
which can be obtained directly 
by using the EBK quantization rules \cite{bt,book}
for the spherically symmetric Hamiltonian.
Note that, before taking the trace integral over the angular
momentum $L$ by the SPM in Eq.~(\ref{poissonsum}), one can formally
consider positive and negative $M$, as those related to the two
opposite directions of motion along a CT (with different signs of the
angular momentum).
They give, of course, equivalent contributions into the trace formula, 
due to a time-reversal symmetry of 
the Hamiltonian, and therefore, one can write simply the
additional factor 2 in
Eq.~(\ref{poissonsum}) but with a further summation over only 
positive integers $M$. It is
in contrast to the standard Poisson summation
trace formula \cite{book} (except for its TF component)
because there is no zero values of the integers in 
Eq.~(\ref{poissonsum}), $n_\varphi/n_r>0$. 
The essential point in the derivations of Eq.~(\ref{poissonsum}) from 
Eq.~(\ref{pstrace}) is that
the generating function $\varPhi_{\rm CT}$
[Eq.~(\ref{phase3})] is independent of the 
angle variables
for {\it families} of the maximal degeneracy $\mathcal{K}=1$
in the integrable Hamiltonian.
Notice that in these derivations, the SPM conditions
were satisfied simultaneously
within the continuum of the stationary points $0 \leq
\varphi$, $\varTheta_r \leq 2 \pi$, which form CTs, but they are not yet
POs generally speaking for arbitrary angular momentum $L$. 
(Exceptions are the cases 
of the complete degeneracy as the spherical HO;
see below.)  The
integration range in Eq.~(\ref{poissonsum}) taken from the minimum,
$L_{-}=0$, to the maximum, $L_{+}$, value (for anticlockwise motion,
for instance) covers the contributions of a whole manifold of closed
and unclosed CTs of the tori in the phase space at the energy surface
around the stationary point, $L=L^*$,
which corresponds to the PO \cite{maf}.  We shall specify the
integration limits $L_{+}$ for the contribution of the
($\mathcal{K}=1$) diameter families $M(n_r=2, n_\varphi=1)$ into
Eq.~(\ref{poissonsum}) in Appendix~\ref{appD}.

Then we apply the stationary phase condition 
with respect to the variable $L$ for the exponent phase $\varPhi_{\rm CT}$
[Eq.~(\ref{phase3})] in the integrand of  
Eq.~(\ref{poissonsum}),
\begin{equation}
\left(\partial \varPhi_{\rm CT}/\partial L\right)^*=0,
\label{statcond1L}
\end{equation}
which is equivalent to the resonance condition (\ref{percond}). 
This condition determines the stationary phase point, $L=L^*=L_{\rm PO}$,
related to the families of the POs $M(n_r,n_\varphi)$.
All these roots of
equation (\ref{percond}) for $\mathcal{K}=1$ families
$M(n_r,n_\varphi)$ are in between the minimum value $L=L^*=0$ for
diameters, and a maximum one $L=L_{C}$, $0 \leq L_{\rm PO} \leq L_{C}$ 
(anticlockwise motion, for
example).  Expanding now the exponent phase $\varPhi_{\rm CT}$ 
[Eq.~(\ref{phase3})] in the variable $L$ up to the second
order, and assuming that there
is no singularities in the curvature (\ref{curv}) for the
contribution of all $\mathcal{K}=1$ families, one has
\begin{equation}
\varPhi_{\rm CT} =
S_{\rm PO}(E) +\frac12 J_{\rm PO}^{(L)}(L- L^*)^2 + \cdots,
\label{phaseexp}
\end{equation}
where $S_{\rm PO}(E)$ is the action along one of the
 isolated PO families determined by Eq.~(\ref{percond}),
\begin{equation}
S_{\rm PO}(E) = 2\pi M \left[n_r\, I_r(E,L^*)+n_\varphi \,L^*\right]\;.
\label{actionpo}
\end{equation}
In this equation, $M$ is the number of repetitions of the primitive
($M=1$) orbit, $I_r(E,L)$ is the energy surface [Eq.~(\ref{actionvar})], 
$L=L^*(n_r,n_\varphi)$ is the solution of the PO
equations (\ref{percond}) or (\ref{statcond1L}). 
 The Jacobian $J_{\rm PO}^{(L)}$ in Eq.~(\ref{phaseexp}) measures the 
stability of the PO with respect to the
variation of the angular momentum $L$ 
at the energy surface,
\begin{gather}
J_{\rm PO}^{(L)} =\left(\frac{\partial^2 S_{\rm CT}}{\partial L^2}
\right)_{L=L^*} =2\pi M n_r K_{\rm PO}, \\
K_{\rm PO} =  \left(\frac{\partial^2 I_r}{\partial L^2}\right)_{L=L_{\rm PO}},
 \label{jacobpar}
\end{gather}
where $K_{\rm PO}$ 
is the curvature (\ref{curv}), (\ref{curvtr}) of the energy 
surface $I_r=I_r(E,L)$ at $L=L^*=L_{\rm PO}$. 

For the sake of simplicity, we shall discuss the simplest 
leading ISPM taking up to the second order
term in the expansion over $(L-L^*)$ for the action phase
[Eq.\ (\ref{phaseexp})], and accounting for only the zeroth 
order component for
the pre-exponential factor in Eq.~(\ref{poissonsum}).  Substituting now
these expansions into Eq.~(\ref{poissonsum}), one can take
the pre-exponential factor off the integral at $L=L^*$.  Thus,
applying Eq.~(\ref{phaseexp}), we are left with the integral over $L$
of a Gaussian type integrand within the finite limits mentioned above
for contributions of the one-parametric polygon-like and
diameter families, including the contribution of boundaries for
$0<n_\varphi/n_r \leq 1/2$.  Taking this integral over $L$ within the
finite limits, one obtains the ISPM trace formula, 
$\delta g^{(\mathcal{K})}(E)$,
for contributions of the one-parametric ($\mathcal{K}=1$) orbits,
\begin{align}
\delta g^{(1)}(E) &=
\Re\sum_{\rm PO} A_{\rm PO}^{(1)}(E) \nonumber \\
& \qquad \times\exp\left[
\frac{i}{\hbar}\,S_{\rm PO}(E)-
i \frac{\pi}{2} \sigma_{\rm PO}-i \phi_{d}\right]\;.
\label{deltagra1}
\end{align}
The sum is taken over the discrete families of the PO
$M(n_r,n_\varphi)$ with $n_r \geq 2\,n_\varphi$, $M \geq 1$ in the 2D
RPL potential, as explained below Eq.~(\ref{phase3}).  
$S_{\rm PO}(E)$ is the action (\ref{actionpo}) along
these POs.  For the amplitudes $A_{\rm PO}^{(1)}$, one finds
\begin{equation}
A_{\rm PO}^{(1)}=\frac{T_{\rm PO}}{\pi\hbar^{3/2}\sqrt{M n_r^3
K_{\rm PO}}}\; \erf(\mathcal{Z}_{\rm PO}^-,\mathcal{Z}_{\rm PO}^+),
\label{ampra1}
\end{equation}
just as for $\mathcal{K}=1$
families in the elliptic billiard \cite{ellipse}, and
the integrable H\'{e}non-Heiles (IHH)
potentials \cite{maf},
with the period $T_{\rm PO}=2\pi n_r/\omega_r=
2\pi n_\varphi/\omega_\varphi$ along the primitive $(n_r,n_\varphi)$
PO.  In the RPL Hamiltonian under consideration, one has
\begin{equation}
T_{\rm PO}=
\frac{\rmd S_{PO}(E)}{\rmd E}=\frac{\pi (\alpha+2)}{\alpha E}\;
\left[n_r I_r\left(E,L_{PO}\right) +n_\varphi L_{PO}\right]
\label{tpora1}
\end{equation}
[see Eqs.~(\ref{actionpo}) for the action $S_{PO}$ and (\ref{scaliolk}) 
with (\ref{scalE}) for the scaling transformations].
In Eq.~(\ref{ampra1}), $K_{\rm PO}$ is the curvature  
of the energy surface $I_r=I_r(E,L)$ [$K_{\rm PO}>0$ at $\alpha > 2$ for RPL
Hamiltonians (\ref{ramodra}); 
see Eqs.~(\ref{curv}), (\ref{percond}) and (\ref{fldef}), 
or Eqs.~(\ref{curvtr})
and (\ref{freq})].
The generalized complex error function 
in Eq.~(\ref{ampra1}) is introduced by
$\erf(u,v) 
=\erf(v)-\erf(u)$ 
with the standard error functions, ${\erf}(z)$, of the
complex arguments $z$. These arguments are specified by
\begin{equation}
\mathcal{Z}_{\rm PO}^{\pm}=\sqrt{-i\pi M n_r K_{\rm PO}/\hbar}\,
(L_{\pm}-L_{\rm PO})\;, 
\label{limra1}
\end{equation}
$L_{-}=0$ and $L_{+}=L_C$ for all $\mathcal{K}=1$ polygon-like PO families
(besides of the diameters, see below). 
For simplicity, the finite integration 
interval of the angular momenta was split
into two parts, $-L_{C} \leq L \leq 0$ and 
$0 \leq L \leq L_C$, where $L_C$ is the
angular momentum of a circle orbit, as mentioned above. 
There are the symmetric stationary points,
$\pm |L^*|$, related to the anticlockwise and clockwise
motions of the particle along the PO in two these phase-space parts. 
As noted above, they give equivalent contributions to the amplitude, 
due to the independence 
of the Hamiltonian of time.  Thus, we have reduced the integration region
to $0 \leq L \leq L_C$, accounting for this time-reversibility symmetry
simply by the factor 2 in Eq.~(\ref{ampra1})
(exceptions are the diameters, for which one has the single
stationary point $L^*=0$, and therefore, 
the time-reversibility degeneracy is one, as it is taken into
account automatically by the limits of the error functions).
For all the polygon-like
and diameter POs
$(n_r\geq 2)$, we also found $L_{-}=0$ for the minimum value of the
angular momentum $L$.

For the Maslov index of the 
considered $\mathcal{K}=1$ PO families 
and the constant phase  $\phi_d$ in Eq.~(\ref{deltagra1}), one obtains
\begin{equation}
\sigma_{\rm PO}^{(1)}=2M n_r,\qquad \phi_d=-\pi/4\;.
\label{maslra1}
\end{equation}
The Maslov index $\sigma_{{}_{\!\rm PO}}$ is determined in terms of the number of
turning and caustic points by the Maslov\&Fedoryuk catastrophe theory, see
Refs.~\cite{fedoryukpr,maslov,maf}.
Note that for the potentials 
with smooth edges, the expression for
the Maslov index $\sigma_{{}_{\!\rm PO}}$ differs from that for
the circular billiard 
\cite{disk,book}.
Note also that the total Maslov phase,  
defined as
a sum of the asymptotic part (\ref{maslra1})
and the argument of the complex density amplitude (\ref{ampra1}),
depends on the energy $E$ and parameter $\alpha$ of
the RPL potential [Eq.~(\ref{ramod}); see Refs.~\cite{ellipse,spheroid}].
This total Maslov phase is changed through the bifurcation points 
smoothly, due
to the phase of the complex error function in the amplitude 
(\ref{ampra1}) in 
Eq.~(\ref{deltagra1}).

For the stationary point $L^*$ far from the ends of the
physical integration
interval, one can extend the integration range to the infinity
from $-\infty$  to $\infty$ (in the case of diameters from zero
to $\infty$).  We then arrive asymptotically at the Berry\&Tabor result
\cite{bt} for the contribution of all $\mathcal{K}=1$
families (\ref{deltagra1}) with the following amplitude:
\begin{equation}
A_{\rm PO}^{(1)}
\rightarrow \frac{d_{\rm PO}T_{\rm PO}}{\pi\hbar^{3/2}\sqrt{
M n_r^3 K_{\rm PO}}}\;,
\label{ampra1ssp}
\end{equation}
where $d_{\rm PO}$ accounts for the discrete degeneracy, $d_{\rm PO}=1$
for diameters $M(2,1)$ $(n_r=2n_\varphi)$, and 2 for all other (polygon-like) POs
$(n_r>2n_\varphi)$ \cite{book}.
In the circular billiard limit ($\alpha\to\infty$), the action
is given by $~S_{\rm PO}(E) \rightarrow p\mathcal{L}_{\rm PO}$ 
with the momentum $p=\sqrt{2mE}$,
and the PO length $\mathcal{L}_{\rm PO}$.
For the curvature $K_{\rm PO}$ [Eqs.~(\ref{curv}) and (\ref{curvtr})], 
one can asymptotically ($\alpha \rightarrow \infty$) obtain
$K_{\rm PO} \rightarrow 1/[\pi pR_0\sin\left(\pi n_\varphi/n_r\right)]\;$.
Substituting all these quantities, $S_{\rm PO}$, $K_{\rm PO}$,
$\sigma_{\rm PO}^{(1)}$ 
[with accounting for
the Maslov-phase contribution of the turning points 
due to the pure reflections 
from the infinite circle walls \cite{disk} as compared to smooth
potentials\cite{maf} in addition to Eq.\ (\ref{maslra1})], 
and the asymptotic amplitude
(\ref{ampra1ssp}) into Eq.~(\ref{deltagra1}), one obtains the well known
trace formula for the circular billiard \cite{disk,book}.
Note that the amplitude
(\ref{ampra1}) of the solution (\ref{deltagra1}) is regular at the
bifurcations which are the boundary points $L=L^*=L_C$ of the action ($L$)
part of the tori as in the elliptic billiard \cite{ellipse}.

Our SSPM result (\ref{ampra1ssp}) coincides with the Berry and Tabor
trace formula \cite{bt}, as adopted to the 2D spherically-symmetric
Hamiltonians by using the simplest expansions of the action phase and
amplitude near the
stationary point (see above), instead of a more general but more
complicated mapping procedure; see more comments in
Ref.~\cite{ellipse}.  The essential difference from the Berry\&Tabor
theory \cite{bt} is that Eq.~(\ref{deltagra1}) covers
all the solutions of the symmetry-breaking problem for the highest
degenerate orbits, such as the one-parametric families in the IHH potential, or
the elliptic and hyperbolic orbits in the elliptic billiard \cite{ellipse}
(see also Refs.~\cite{maf,migdal}).  Within the SPM of the extended
Gutzwiller approach \cite{strm,maf,migdal}, we have to derive
separately the contributions of the other orbits as the circle
$\mathcal{K}=0$ POs in the RPL potentials beyond the semiclassical
Poisson summation-like trace formula (\ref{poissonsum}) (with the
restrictions to the range of the $n_r$ and $n_\varphi$ integer
variables).  We emphasize that the ISPM trace formula (\ref{deltagra1}) for
the one-parametric families contains the end
contributions related to the finite limits of integrations in the
error functions.  However, this trace formula can be only applied to
the contribution of such
families, as pointed out above in its derivation from the trace
formula (\ref{pstrace}).
Therefore, there is no contributions of the circle
orbits in Eqs.~(\ref{poissonsum}) and (\ref{deltagra1}).
As shown below, these orbits correspond to the separate contribution of 
the {\it isolated ($\mathcal{K}=0$) 
stationary-phase} point $L^*=L_{C}$ (as for the IHH potential \cite{maf}, 
for example).

\subsection{Circle orbits ($\mathcal{K}=0$)}
\label{sec3B}

In contrast to the derivations of contributions of the orbits
with the highest degeneracy $\mathcal{K}=1$, 
we now take into account the
existence of the isolated stationary point of the action phase 
$\varPhi_{\rm CT}$ (\ref{legendtrans}) in the radial spherical 
phase-space variables
$r^{\prime\;*}=r^{\prime\prime\;*}=r_{{}_{\!C}}, \,\,
p_r^{\prime\;*}=p_r^{\prime\prime\;*}=0~$. 
After the transformation
of the integration variables in Eq.~(\ref{pstrace}) to the spherical
phase space coordinates $\{r^\prime,\varphi^\prime;
p_r^{\prime\prime},L \}$, it is convenient first to perform the exact
integrations over $L$ by using the energy conserving $\delta$-function,
and over the cyclic azimuthal angle $\varphi'$ leading simply to $2
\pi$ as above 
($\int \rmd \varphi'/\omega_\varphi=T_{\varphi,{\rm CT}}$ is the 
primitive rotation period).
Thus, one finds
\begin{align}
&g_{\rm scl}(E)
=\frac{2}{(2\pi\hbar)^2}\, \Re\sum_{\rm CT}
\,\int \rmd r^\prime
\,\int \rmd p_r^{\prime\prime}\,T_{\varphi,{\rm CT}} \nonumber \\
& \times \left|\mathcal{J}_{\rm CT}(p^\prime_r,p^{\prime\prime}_r)\right|^{1/2}\; 
 \exp\left[\frac{i}{\hbar}
\varPhi_{\rm CT} - i \frac{\pi}{2}
\mu_{{}_{\!\rm CT}} - i \phi_d\right]\;.
\label{pstracerpC}
\end{align}
The additional factor 2 accounts for the equivalent contributions of two CTs 
for the particle motion in the two opposite directions 
(with the opposite signs of the angular momentum as above).
The stationary phase condition for the SPM integration over 
the radial momentum $p_r^{\prime\prime}$ in Eq.~(\ref{pstracerpC})
is written as
\begin{equation}
\left(\frac{\partial \varPhi_{\rm CT}}{\partial p_r^{\prime\prime}}\right)^*
\equiv \left(r^{\prime} -r^{\prime\prime}\right)^*=0\;.
\label{statcondCp}
\end{equation}
The solution of this equation is the isolated 
stationary point $p_r^{\prime\prime}=
p_r^{\prime\prime\;*}=p_r^*=0$.
The phase $\varPhi_{\rm CT}$ [Eq.~(\ref{legendtrans})] is expanded
in the momentum $p_r^{\prime\prime}$  near this point $p_r^{\prime\prime\;*}=0$
in power series,
\begin{equation}
\varPhi_{\rm CT}=\varPhi_{\rm CT}^* + \frac{1}{2}\mathcal{J}_{\rm CT}^{(p)}\,
\left(p_r^{\prime\prime}-p_r^*\right)^2+\cdots, 
\label{phipiexpC}
\end{equation}
where the Jacobian is given by
\begin{equation}
\mathcal{J}_{\rm CT}^{(p)}=
\left(\frac{\partial^2 \varPhi_{\rm CT}}{\partial
  p_r^{\prime\prime\,2}}\right)^*=
\left[\frac{2 \pi M n_r K}{
\left(\partial p_r^{\prime\prime}/\partial L\right)^2}\right]^*\;.
\label{jacpC} 
\end{equation}
The star implies again that the
corresponding quantity is taken at the stationary point, 
$p_r^{\prime\prime}=p_r^{\prime\prime\;*}=0$.
Using the 2nd order expansion of the exponent phase
(\ref{phipiexpC}) 
and taking the pre-exponent amplitude factor 
off the integral at this stationary point,
one gets the internal integral
over $p_r^{\prime\prime}$ in Eq.~(\ref{pstracerpC}) 
in terms of the error function as 
in the previous section.  According to
Eq.~(\ref{legendtrans}), 
with the radial-coordinate closing condition (\ref{statcondCp}) 
for the CTs,
the short phase $\varPhi_{\rm CT}^*$ in 
Eq.~(\ref{phipiexpC}) 
can be written in terms of the corresponding variables as 
$\varPhi_{\rm CT}^* = \int_{r'}^{r''} p_r\,\rmd r$.
Taking then into account the CT closing condition (\ref{statcondCp}), 
$r^\prime=r^{\prime\prime}=r$, for
the stationary phase equation in the integration over the radial
$r$ coordinate perpendicular to the circle orbit, 
one results in
\begin{equation}
\left(\frac{\partial \varPhi_{\rm CT}^{*}}{\partial r^{\prime\prime}}
+\frac{\partial \varPhi_{\rm CT}^{*}}{\partial r'}\right)^{*} \equiv
\left(p_r^{\prime\prime}-p_r^\prime\right)^{*}=0\;.
\label{statcondCr}
\end{equation}
Therefore, together with Eq.~(\ref{statcondCp}), one has the PO conditions
related to the circular orbit $r=r^*=r_{{}_{\!C}}$ and $L=L^*=L_C$ 
(see Appendix~\ref{appA}). 
As usually within the SPM, we expand now the phase $\varPhi_{\rm CT}^*$ 
in the radial
coordinate $r$ near this $r^*=r_{{}_{\!C}}$,
\begin{equation}
\varPhi_{\rm CT}^*=MS_{C}+
\frac{1}{2}\mathcal{J}_{MC}^{(r)}\,
\left(r-r_C\right)^2 +\cdots, 
\label{phirexp}
\end{equation}
where $S_{C}$ is the action along the primitive circle PO ($C$),
\begin{equation}
\mathcal{J}_{MC}^{(r)} 
=\left(-\frac{\partial p_r'}{\partial
  r^{\prime}} - 2\frac{\partial p_r^\prime}{\partial r^{\prime\prime}}
+ \frac{\partial p_r^{\prime\prime}}{\partial
  r^{\prime\prime}}\right)_{MC}^*\;.
\label{jacrC} 
\end{equation}
Again, using the action phase expansion (\ref{phirexp}) 
at the second order as the 
simplest ISPM approximation, and taking the pre-exponent amplitude 
factor at the isolated stationary
point $r=r_C$ off the integral, one finally obtains
\begin{align}
\delta g_{\{MC\}}^{(0)}(E)
&=\Re \sum_{M=1}^\infty \,A_{MC}^{(0)}(E)\, \nonumber \\ &\times
\exp\left[\frac{i}{\hbar}\,MS_{C}(E) - i \frac{\pi}{2}
\sigma^{(0)}_{MC}-i \phi^{(0)}_{d}\right]\;.
\label{deltagra0}
\end{align} 
The sum runs all repetitions of the circle orbit $MC$ with
$M=1,2,\cdots$ being positive integers. The time-reversal symmetry of the 
Hamiltonian (equivalence of the contributions of
both angular momenta and repetition numbers 
with opposite signs)
was taken into account by the factor 2 in 
Eq.~(\ref{pstracerpC}).
The action $S_C(E)$ along the primitive $C$ orbit is given by
\begin{equation}
S_C(E)=\oint_{C} p_\varphi \,\rmd\,\varphi = 2\pi\,L_C
\label{actionC}
\end{equation}
with $L_C$ shown explicitly in Eq.~(\ref{rcLcd2Fcra}). 
In Eq.~(\ref{deltagra0}),
$\sigma^{(0)}_{MC}$ is the
Maslov index determined by the number of caustic and turning points
along the circle orbit,
according to the Fedoryuk\& Maslov catastrophe
theory\cite{fedoryukpr,maslov,maf,maf}, 
\begin{equation}
\sigma^{(0)}_{MC}=4M,\qquad \phi^{(0)}_d=0\; .
\label{maslra0}
\end{equation}
For the amplitudes $A_{MC}^{(0)}(E)$ in Eq.~(\ref{deltagra0}), one finds
\begin{equation}
A_{MC}^{(0)}=\frac{T_C}{4\pi\hbar\sqrt{F_{MC}}}\;
\erf(\mathcal{Z}_{p,MC}^{(-)},\mathcal{Z}_{p,MC}^{(+)})\;
\erf(\mathcal{Z}_{r,MC}^{(-)},\mathcal{Z}_{r,MC}^{(+)})\;,
\label{ampra0}
\end{equation}
where $T_C$ is the period of the primitive ($M=1$) orbit $C$,
\begin{equation}
T_C=\frac{\rmd S_C(E)}{\rmd E}=\pi L_C\;\frac{\alpha+2}{\alpha E}\;;
\label{periodc}
\end{equation}
see Eqs.~(\ref{actionC}), (\ref{scaliolk}) and (\ref{scalE}).
In Eq.~(\ref{ampra0}), $F_{MC}$ is the Gutzwiller stability factor 
\cite{gutz} of the circle orbits [Eq.~(\ref{fgutzC})].
The arguments of the error functions 
in Eq.~(\ref{ampra0}) can be 
transformed to the following invariant form (see Appendix~\ref{appE}): 
\begin{eqnarray}
\mathcal{Z}_{p,MC}^{(\pm)}&=&\sqrt{-\frac{i}{\hbar}\,
\pi\,M\,\sqrt{\alpha+2}\,K_C}
\,\left(L_{\pm}-L_C\right), \nonumber\\
&&\qquad L_{+}=L_C, \quad L_{-}= 0\;, \nonumber\\  
\mathcal{Z}_{r,MC}^{(\pm)} &=& \sqrt{\frac{i\, F_{MC}}
{4 \pi\,M\, \hbar\,(\alpha+2)^{3/2}\;K_C}}\;
\varTheta_r^{(\pm)}\;, \nonumber\\
&&\qquad \varTheta_r^{(+)}=2\pi,\qquad \varTheta_r^{(-)}=0\;.
\label{limra0}
\end{eqnarray}
Here, $L_{\pm}$ are the maximum and minimum 
values of the angular-momentum integration variable for 
the contribution of the circle orbits, 
$K_C$ is their curvature (see Appendix~\ref{appE}),
\begin{equation}
K_C = \frac{(\alpha+1)(\alpha-2)}{12 \,(\sqrt{\alpha+2})^3\,
  L_C}\;. 
\label{curvraC}
\end{equation}
The simplest approximation in Eq.~(\ref{limra0}) is $L_+=L_C$, $L_-=0$; and
$\varTheta_r^{-}=0$, $\varTheta_r^{(+)}=2\pi$,
which correspond to the total physical phase space accessible for the
classical motion. 
The factors $\sqrt{\alpha+2}$ in front of the curvature $K_C$
 appear because of the frequency ratio 
$f(L)=\omega_\varphi/\omega_r$  
for the circle orbits for any parameter $\alpha \geq 2$; 
see Eqs.~(\ref{fldef}), (\ref{omtcra}) and (\ref{omc}). 
For $\alpha=4$; the
period $T_C$ [Eq.~(\ref{periodc})], action $S_C$ [Eq.~(\ref{actionC})],
 curvature $K_C$ [Eq.~(\ref{curvraC})], and stability factor
$F_{MC}$ [Eq.~(\ref{fgutzC})] for the circle orbits are identical
to those obtained in Ref.~\cite{trap}.
We used also the properties of the 
Jacobians for transformations
of the different coordinates, in particular, 
given by Eq.~(\ref{stabfactjac}).
Note that  after applying the stationary phase
conditions $r^*=r_{{}_{\!C}}$ [Eq.~\ref{statcondCp}] and $p_r^*=0$
[Eq.~(\ref{statcondCr})], the angular momentum $L$ of the circular orbits
as function of the $r$ and $p_r$ becomes the isolated stationary point $L^*=L_C$
at the boundary of the classically accessible phase space.
Notice also that the asymptotic
Maslov phase is defined traditionally in terms of the
Maslov index $\sigma^{(0)}_{MC}$ [Eq.~(\ref{maslra1})].
There is again the two components of the Maslov phase in the ISPM
trace formula (\ref{deltagra0}) for the $MC$ orbits.  
One of them is the
asymptotic constant part (\ref{maslra1})  independent of the 
energy.  Another part is the argument of the complex 
amplitudes $A_{MC}^{(0)}$ [Eq.~(\ref{ampra0})], that changes 
continuously through the bifurcation points.
The total Maslov phase  for the 
circle POs is given by the sum of these two contributions,  
which ensures a smooth transition of the trace formula
(\ref{deltagra0}) for the contribution of the circle POs through
the bifurcation points.

In the asymptotic limit of the non-zero integration 
boundaries, $L_{-}\rightarrow  -\infty$ and 
$\varTheta_r^{+} \rightarrow \infty$, i.e., far from any 
bifurcations
$\alpha_{\rm bif}$ [Eq.~(\ref{bifeqra}), including the HO symmetry breaking
at $\alpha=2$], 
the expression (\ref{ampra0}) tends (through the Fresnel functions
of the corresponding 
real positive arguments) to the amplitude of the Gutzwiller
trace formula for isolated orbits \cite{gutz,book},
\begin{equation} 
A_{MC}^{(0)}(E)
\rightarrow  \frac{1}{4\pi \,\hbar}\, 
\frac{T_{C}}{\sqrt{F_{MC}}}\;.
\label{ampra0ssp}
\end{equation} 
In this 
limit, the asymptotic
Maslov index $\sigma^{(0)}_{MC}$ and 
$\phi_d^{(0)}$ in Eq.~(\ref{deltagra1}) are given by
Eq.~(\ref{maslra0}). Notice that the number coefficient in 
Eq.~(\ref{ampra0ssp}) differs from the SSPM Gutzwiller's expression (5.36) 
of Ref.~\cite{book} by factor 1/4. The reason is that 
the two stationary-phase
points $r^{\prime\;*}=r_C$ and $p_r^{\prime\prime\;*}=0$ belong to the boundary
of the physical  $\{r^{\prime}, p_r^{\prime\prime}\}$ phase-space 
integration volume in Eq.~(\ref{pstracerpC}); while 
in Ref.~\cite{gutz}, all the stationary points are assumed to be 
internal ones which are far away from the integration boundary. 
Eq.~(\ref{ampra0ssp}) can be derived directly
from Eq.~(\ref{pstracerpC}) by using the SSPM. To realize this
within the SSPM,
one may extend in Eq.~(\ref{pstracerpC}) 
the $r'$ integration range from $\{r'=0,r_{{}_{\!C}}\}$ to 
$\{-\infty,r_{{}_{\!C}}\}$, and similarly,  
the $p_r^{\prime\prime}$ integration one to  $\{0,\infty\}$,
assuming that the lower $r'$ and upper $p_r^{\prime\prime}$ 
integration limits are far away from the 
corresponding other (stationary-point) integration boundaries. 

For the opposite limit  to the bifurcations ($F_{MC} \rightarrow 0$,
when $\alpha \rightarrow \alpha_{\rm bif}$),
one finds that the both arguments of the second error function
in Eq.~(\ref{ampra0}) 
tend to zero as $\sqrt{|F_{MC}|}$, see Eq.~(\ref{limra0}).
 The  Gutzwiller stability 
factor $F_{MC}$, going to zero, is exactly canceled by the same one in 
the denominator, and we arrive at
\begin{align}
A_{MC}^{(0)}(E) \rightarrow &
\frac{T_{C}}{4\,\hbar^{3/2}\sqrt{\pi M\,(\alpha+2)^{3/2}\,K_{C}}}
\nonumber \\ &\qquad
\times\erf\left(\mathcal{Z}_{p,\;MC}^{(-)},\mathcal{Z}_{p,\;MC}^{(+)}\right)\; 
e^{i \pi/4}\;.
\label{amp1ispbif}  
\end{align}
Thus, in contrast to the SSPM divergences,
one obtains the finite results at 
the bifurcations within the ISPM. 
Notice that the 
enhancement in order of $\hbar^{-1/2}$ with respect to the Gutzwiller
asymptotic amplitude (\ref{ampra0ssp}) takes place 
locally near the bifurcation
points.
Note also that at the circular billiard limit, when
$K_{C} \rightarrow \infty$ (separatrix), one finds a continuous
limit which is zero in the case of the RPL potential.

\subsection{Total trace formula for the oscillating level density}
\label{tottraceesc}

The total semiclassical oscillating (shell) correction to the 
level density (\ref{pstrace}) for the RPL potentials
in two dimensions is thus given by
\begin{equation}
\delta g_{\rm scl}(E)=\delta g_{\rm scl}^{(1)}(E)
+\delta g_{\rm scl}^{(0)}(E)\;,
\label{totdensc}
\end{equation}
where
\begin{align}
\delta g_{\rm scl}^{(\mathcal{K})}(E)
&=\Re\sum_{\rm PO} A_{\rm PO}^{(\mathcal{K})}(E)\; \nonumber \\
&\times
\exp\left[\frac{i}{\hbar}S_{\rm PO}(E)-i\frac{\pi}{2}\sigma^{(\mathcal{K})}_{\rm PO}
-i\phi_d^{(\mathcal{K})}\right]\;.
\label{dgrake}
\end{align}
The amplitudes $A_{\rm PO}^{(\mathcal{K})}$ 
[see Eqs.~(\ref{ampra1}) for $\mathcal{K}=1$ and
 (\ref{ampra0}) for $\mathcal{K}=0$], actions $S_{\rm PO}$, Maslov 
indexes $\sigma^{(\mathcal{K})}_{\rm PO}$, and constant phases
 $\phi^{(\mathcal{K})}_d$ [Eqs.~(\ref{maslra1}) 
and (\ref{maslra0})] 
were specified above.

Using the scale invariance (\ref{scaling}), one may factorize 
the action integral
\begin{equation}
S_{\rm PO}(E) 
=\left(\frac{E}{E_0}\right)^{\frac12+\frac{1}{\alpha}}
\oint_{{\rm PO}(E=E_0)}\bfp\cdot \rmd \bfr \nonumber \\
\equiv\varepsilon\tau_{{}_{\!\rm PO}}\;.
\label{actionsc}
\end{equation}
In the last equation, we define the scaled energy
$\varepsilon$ and scaled period $\tau_{{}_{\!\rm PO}}$ by
\begin{equation}
\varepsilon=\left(\frac{E}{E_0}\right)^{\frac12+\frac{1}{\alpha}},
\quad
\tau_{{}_{\!\rm PO}}=\oint_{{\rm PO}(E=E_0)}
\bfp \cdot \rmd \bfr\;.
\label{eq:scaled_E_and_T}
\end{equation}
To realize the advantage of the scaling
invariance (\ref{scaling}), it is helpful to use the scaled
energy (period) in place of the corresponding 
original variables.
For the  HO, one has $\alpha=2$, and the 
scaled energy and period are proportional to the unscaled quantities.
For the cavity potential ($\alpha \rightarrow \infty$), they are proportional 
to the momentum $p$ and length $\mathcal{L}_{\rm PO}$, respectively.

Using the transformation of the energy $E$ to the 
scaled energy $\varepsilon$,
one can introduce the dimensionless scaled-energy level density.
The advantage of this transformation is that a nice
plateau condition is always found
in the Strutinsky SCM smoothing procedure 
by using the scaled spectrum $\varepsilon_i$
(see Refs.~\cite{ellipse,spheroid} for the case of the billiard limit
$\alpha\rightarrow\infty$).
Then, one can use a simple relation between the original and
scaled-energy level densities, 
\begin{equation}
\mathcal{G}(\varepsilon)=\sum_i\delta(\varepsilon-\varepsilon_i)
=g(E)\frac{\rmd E}{\rmd\varepsilon}\;.
\label{denscaledef}
\end{equation}
For the semiclassical
oscillating part of the level density (\ref{denscaledef}), one finds
\begin{align}
&\delta\mathcal{G}_{\rm scl}^{(\mathcal{K})}(\varepsilon)
=\frac{\rmd E}{\rmd\varepsilon}\delta g^{(\mathcal{K})}(E)
=\sum_{\rm PO}\delta\mathcal{G}_{\rm PO}^{(\mathcal{K})}(\varepsilon)
 \nonumber\\ &\qquad
=\Re\sum_{\rm PO}\mathcal{A}_{\rm PO}^{(\mathcal{K})}(\varepsilon)
 \exp\left[\frac{i}{\hbar}\varepsilon\tau_{{}_{\!\rm PO}}
 -\frac{i\pi}{2}\sigma^{(\mathcal{K})}_{\rm PO}
 -i\phi^{(\mathcal{K})}_d\right]\,, \nonumber \\
&\qquad \mathcal{A}_{\rm PO}^{(\mathcal{K})}(\varepsilon)
 =\frac{\rmd E}{\rmd\varepsilon} A_{\rm PO}^{(\mathcal{K})}(E).
\label{dgrakeps}
\end{align}
The simple form of the phase function (\ref{actionsc}) 
enables us also to make easy use of 
the Fourier transformation technique.  The Fourier transform of the
semiclassical scaled-energy level density with respect to the scaled
period  $\tau$ is given by
\begin{equation}
F(\tau)=\int \rmd\varepsilon\; \mathcal{G}(\varepsilon)e^{i\varepsilon\tau/\hbar}
\approx F_0(\tau)+\sum_{\rm PO}\widetilde{\mathcal{A}}_{\rm PO}
\delta(\tau-\tau_{{}_{\!\rm PO}})\;,
\label{fourier_power}
\end{equation}
which exhibits peaks at periodic orbits $\tau=\tau_{{}_{\!\rm PO}}$. $F_0(\tau)$ 
represents the Fourier transform of the smooth Thomas-Fermi
level density and 
has a peak at $\tau=0$ related to the zero-action trajectory 
\cite{migdal}. 
Thus, from the Fourier transform of the scaled-energy quantum-mechanical 
level density (\ref{denscaledef}),
\begin{equation}
F(\tau)
=\sum_i e^{i\varepsilon_{i}\tau/\hbar}, \qquad 
\varepsilon_{i}=\left(\frac{E_i}{E_0}\right)^{\frac12+\frac{1}{\alpha}},
\label{fourier_power_qm}
\end{equation}
one can directly extract the information about classical PO
contributions.  The trace formula (\ref{totdensc}) has the correct
asymptotic SSPM limits to the Berry\&Tabor results
(\ref{deltagra1}), (\ref{ampra1ssp}) for $\mathcal{K}=1$ polygon-like
(including the diameters) and to the Gutzwiller trace formula
(\ref{deltagra0}), (\ref{ampra0ssp}) for $\mathcal{K}=0$ circle POs.
As shown in the sections \ref{sec3A} and \ref{sec3B}, one obtains also 
the limit of the trace formula
[Eqs.~(\ref{totdensc}) and (\ref{dgrake})] to that of the circular billiard 
$\alpha \rightarrow \infty$ \cite{disk,book}. In this limit
one has obviously zero for the circle orbit contributions 
as for the potential barrier separatrix in the IHH potential \cite{maf}.

For comparison with the quantum level densities obtained by the SCM, 
we need also to perform a local averaging of the trace
formula  (\ref{totdensc}) over the spectrum. 
As this trace formula is given through
the sum of the individual PO terms everywhere (including
the bifurcation regions), one can approximately take 
the folding integrals
over energies  
in terms of the Gaussian weight factors with a width
parameter $\varGamma \ll E_F$ .   As the result, one obtains the 
Gaussian-averaged oscillating level density in the analytical
form \cite{strm,book,migdal}:
\begin{equation}
\delta g_{{}_{\!\varGamma}}(E)=\sum_{\rm PO} \delta g_{{}_{\!\rm PO}}(E) \,
\exp\left[-\left(t_{{}_{\!\rm PO}} \varGamma/\hbar\right)^2\right]\;.
\label{avdeltadentot}
\end{equation}
Adding the TF smooth component
$g_{{}_{\!\rm TF}}(E)$ \cite{book} to this oscillating component, 
one results in the total trace formula:
\begin{equation}
g_{{}_{\!\Gamma}}(E)=g_{{}_{\!\rm TF}}(E) + \delta g_{{}_{\!\Gamma}}(E),
\label{denstot}
\end{equation}
where
\begin{align}
g_{{}_{\!\rm TF}}(E) &= \frac{1}{(2 \pi \hbar)^2}\,\int \rmd\bfr  
\int \rmd\bfp \,\,\delta \left(E-\frac{p^2}{2m}-V(r)\right) \nonumber \\ &=
 \frac{m r^2_{\rm max}}{2\hbar^2}=
\frac{1}{2E_0}\left(\frac{E}{E_0}\right)^{2/\alpha}\;. 
\label{TFden}
\end{align}  
Here,
$r_{\rm max}$ is the maximal turning point 
(one of solutions
of the equation $V(r)=E$), which is given by
$r_{\rm max}=R_0 (E/E_0)^{1/\alpha}$ for the RPL Hamiltonian
(\ref{ramodra}), and we put
$E_0=\hbar^2/mR_0^2$ in the last expression of Eq.\ (\ref{TFden}).

Using the scaled-energy transformation (\ref{denscaledef}) of the 
oscillating part (\ref{avdeltadentot}) of the 
Gaussian-averaged level density
[Eq.\ (\ref{denstot})], 
one finally obtains the semiclassical scaled-energy trace formula:
\begin{align}
\delta\mathcal{G}_\gamma(\varepsilon)&=\sum_{\mathcal{K}=0}^1  
\delta\mathcal{G}_\gamma^{(\mathcal{K})}(\varepsilon)
\nonumber \\ &=
\sum_{\mathcal{K}=0}^1 \sum_{\rm PO} 
\delta\mathcal{G}_{\rm PO}^{(\mathcal{K})}(\varepsilon)\,
\exp{
\left[-\left(\frac{\tau_{{}_{\!\rm PO}} \gamma}{2\hbar}\right)^2\right]}\;,
\label{avdenra}
\end{align}
Here, $\delta\mathcal{G}_{\rm PO}^{(\mathcal{K})}(\varepsilon)$ is
given by Eq.~(\ref{dgrakeps}),  $\gamma$ is a dimensionless
width parameter
used for the Gaussian averaging over the scaled 
spectrum $\varepsilon_i$. For the scaled-energy 
Thomas-Fermi density component, one finds
\begin{equation}
\mathcal{G}_{\rm TF}(\varepsilon)=
g_{{}_{\!\rm TF}}(E)\frac{\rmd E}{\rmd\varepsilon}
=\frac{\alpha}{2+\alpha}\,\varepsilon\;.
\label{scldenstyraTF}
\end{equation}

\subsection{The shell correction energies}
\label{esc}

The semiclassical PO shell correction energies
$\delta U_{\rm scl}$ is given by 
\cite{strm,ellipse,spheroid,book,migdal}
\begin{equation}
\delta U_{\rm scl}
 = 2 \sum_{\rm PO} \frac{\hbar^2}{t_{{}_{\rm PO}}^2}\,\delta g_{{}_{\rm PO}}(E_F), 
\label{escscl}
\end{equation}
where $t_{{}_{\rm PO}}=MT_{\rm PO}(E_F)$ is the period of particle motion
along the PO (taking into
account its repetition number $M$) at the Fermi energy $E=E_F$.
The Fermi energy $E_F$ as function of the particle number $N$
is determined by the particle number conservation,
\begin{equation}
N=2 \sum_i n_i=2\int_0^{E_F} \rmd E\,g(E)\;, 
\label{partnum}
\end{equation}
where $n_i=\theta(E_F-E_i)$ are the occupation numbers.
The factors 2 in Eqs.~(\ref{escscl}) 
and (\ref{partnum}) account for the spin 
degeneracy of Fermi particles with spin 1/2.

Note that the shell correction energies $\delta U$ which are
the observed physical quantities do not
contain an arbitrary averaging parameter $\varGamma$, in contrast to the
level density $g_{{}_{\!\varGamma}}(E)$.  The convergence of the PO
sum (\ref{escscl}) to shorter POs (if they occupy enough large
phase-space volume) is ensured by the additional factor in front
of the oscillating density components $\delta g_{{}_{\!\rm PO}}$ which is inversely
proportional to square of the PO period $t_{{}_{\!\rm PO}}$.

In the quantum SCM calculations, the shell correction energies are
usually obtained by extracting the oscillating part from a sum of the
single-particle energies,
Note that the direct application
of the SCM average procedure to the spectra $E_i$ of RPL 
potentials (except for the HO limit) does not give any good
plateau condition as for the level density $g(E)$
in Eq.~(\ref{denscaledef}).
However, one may find rather a good plateau in 
the SCM application
to a sum of the single-particle scaled energies $\varepsilon_i\;$,
$\mathcal{U}=2 \sum_{i} n_i\varepsilon_i\;$. 
Applying exactly the same derivations 
of Eq.~(\ref{escscl}) to
the semiclassical trace formula for
the oscillating part of $\mathcal{U}$, 
one gets
\begin{equation}
\delta\mathcal{U}_{\rm scl}
= 2 \sum_{\rm PO}\frac{\hbar^2}{\tau_{\rm PO}^2}\delta\mathcal{G}_{\rm PO}
 (\varepsilon_{{}_{\!F}})\,. \label{escscl2}
\end{equation}
Here, the scaled Fermi energy $\varepsilon_{{}_{\!F}}$ is determined by
\begin{equation}\label{partnum2}
N=2\int_0^{\varepsilon_{{}_{\!F}}}\mathcal{G}(\varepsilon) 
\rmd \varepsilon\;.
\end{equation}
Using now the obvious relations
$~t_{{}_{\!\rm PO}}=\tau_{{}_{\!\rm PO}}\;\rmd \varepsilon/\rmd E~$ 
and 
$~\delta g_{{}_{\!\rm PO}}(E)=\delta\mathcal{G}_{\rm PO}(\varepsilon)\;
\rmd \varepsilon/\rmd E~$
in Eq.~(\ref{escscl}), one obtains
\begin{equation}
\delta U_{\rm scl}
=\left(\frac{\rmd E}{\rmd\varepsilon}\right)_{\varepsilon_{{}_{\!F}}}
 \delta\mathcal{U}_{\rm scl}\;. \label{enra1ra}
\end{equation}
Thus, we arrive at the simple relation between the
original shell-correction energy $\delta U$ [Eq.~(\ref{escscl})] 
and the scaled one $\delta  \mathcal{U}$, valid for both semiclassical
and quantum (neglecting the second order terms in the shell fluctuations
of the Fermi energy) calculations:
\begin{equation}
\delta U
=\left(\frac{\rmd E}{\rmd\varepsilon}\right)_{\varepsilon_{{}_{\!F}}}
 \delta\mathcal{U}
=E_0\frac{2 \alpha}{\alpha+2}\varepsilon_F^{(\alpha-2)/(\alpha+2)}
 \delta \mathcal{U}. \label{escvsu}
\end{equation}
This relation can be also directly obtained 
by using the standard quantum SCM relations
of the first-order shell-correction energy
$\delta U$ to  the oscillating part of the level density 
$\delta g(E)$ up to the same second order
terms in the Fermi energy oscillations [2],
and corresponding ones for the scaled quantities,
\begin{equation}
\delta \mathcal{U} = 
2 \sum_i \delta n_i \varepsilon_i =
2 \int_0^{\varepsilon_{{}_{\!F}}} \rmd \varepsilon \;
 \left(\varepsilon-\varepsilon_{{}_{\!F}}\right)\; 
\delta \mathcal{G}(\varepsilon)\;.
\label{qmrelscm2}
\end{equation}
In these derivations, $\delta n_i=n_i-{\widetilde n}_i$
represents
the oscillating part of the occupation number defined by
subtracting the smooth part $\widetilde{n}_i$ from the exact one.
We applied also 
the usual transformations 
from the Fermi energies to the particle 
numbers by using Eqs.~(\ref{partnum}) and (\ref{partnum2}), as well as the 
definitions of the averaged
Fermi energy ${\widetilde E}_F$, and the scaled one 
${\widetilde \varepsilon}_F$,
\begin{equation}
N=2 \int_0^{{\widetilde E}_F} \rmd E \; \widetilde{g}(E)=
2 \int_0^{{\widetilde \varepsilon}^{}_F} 
\rmd \varepsilon \; \widetilde{\mathcal{G}}(\varepsilon)\;.
\label{smoothef}
\end{equation}

\subsection{Harmonic oscillator limit}
\label{sec:holimit}

In the isotropic harmonic oscillator limit [$\alpha \rightarrow 2$  
in the power-law potential (\ref{ramod})], the energy surface
is simplified to the linear function in actions,
\begin{equation}
E=\omega_{r}\,I_{r} + \omega_{\varphi}\,I_{\varphi} =
\omega_{\varphi}\,\left(2\,I_{r} + L\right)\;.
\label{Eho}
\end{equation}
Therefore, in this limit
the curvature $K_{\rm PO}$  
for all POs [including the maximum value 
$L=L_{C}=E/\omega_\varphi$ for the circle  orbits, Eq.~(\ref{curvraC}), 
and $L=0$ 
for diameter ones, Eq.~(\ref{curvraD})]
and  stability factor  
$F_{MC}$ [Eq.~(\ref{fgutzC})] for the $MC$
 orbits 
turn into zero.  However, there is no
singularities in the ISPM trace formulas (\ref{deltagra1}) for the 
contributions of all $\mathcal{K}=1$ families  and 
(\ref{deltagra0}) for the circle orbits in the limits  
$K_{\rm PO} \rightarrow 0$ and $F_{MC} \rightarrow 0$.
The arguments of both error functions, $\propto \sqrt{K_{C}}$ and
$\propto \sqrt{F_{MC}/K_C}$  
in Eq.~(\ref{ampra0}), for instance, approach zero and singularities 
are canceled with the same ones in the
denominators of the multipliers
in front of them, and similarly, in Eq.~(\ref{deltagra1}) for one 
error function; see
Eqs.~(\ref{ampra1}), (\ref{limra1}), (\ref{ampra0}) and  (\ref{limra0}) 
with the help of Eq.~(\ref{amp1ispbif}).  Therefore, 
one has a continuous limit of the total trace 
formula (\ref{totdensc}) for $\alpha \rightarrow 2$.
Moreover, in this limit, 
one obtains exactly the same half of the HO trace formula for 
the $MC$
orbit contribution (\ref{deltagra0}) and
the $M(2,1)$ diameter one [Eq.~(\ref{deltagra1})]
up to the relatively small higher-order corrections in $\hbar$ [see   
also Eq.~(3.68) of Sec.~3.2.4 in Ref.~\cite{book}], 
\begin{equation}
 g_{\{MC\}}^{(0)}(E) \rightarrow \frac{1}{2}\,
 \delta g_{\rm HO}^{(2)}(E),\quad
g_{\{MD\}}^{(1)}(E) \rightarrow \frac{1}{2}\, \delta g_{HO}^{(2)}(E)\;.
\label{holimC}
\end{equation}
Here, $\{MC\}$ and $\{MD\}$ represent sum of all repetitions of
circle and diameter orbits, $M=1,2,...$, respectively.
Thus, the HO limit of the sum of the circle and diameter orbit contributions
into the (averaged) level density and the energy 
shell corrections is exactly 
analytically given by the corresponding HO trace formulas.
We point out that for $\alpha \to 2$,
the contributions of circle $MC$ and diameter 
$M(2,1)$ orbits
encounter local increases of the degeneracies $\mathcal{K}$
by 2 and 1 units, respectively.

As noted above, in the HO limit $\alpha \rightarrow 2$, only the 
diameter $M(2,1)$
and the circle $MC$ (both with repetitions) survive, 
and they form $\mathcal{K}=2$ families in the HO potential.
Taking into account also that the angular momentum for the diameters
is always zero, $L^*=0$, and for the circle orbits 
$L^*= L_C$, we shall assume that 
the integration over $L$ for the diameters is performed from $L_{-}=0$
to $L_{+}=L_C/2$ and for the circle orbits from $L_{-}=0$
to $L_{+}=L_C$, such that they give naturally equivalent contributions
into the HO trace formula, as shown in Eq.~(\ref{holimC}), see
also Ref.~\cite{maf}. 
The difference is in the integration limits for the circle
orbits [Eq.~(\ref{limra0})], in contrast to
Eqs.~(\ref{limra1}) and (\ref{bcoefD}) for the diameter boundaries. 
Notice that the contribution of the polygon-like one-parametric orbits, 
$\delta g^{(1)}(E)$, disappears in the ghost HO limit.
Thus, one obtains the continuous
transition of the oscillating part of the ISPM level density
$\delta g_{\rm scl}(E)$
through all bifurcation points,
including the HO symmetry breaking.

\section{Amplitude enhancement and 
comparison with quantum results}
\label{sec4}

\begin{figure}
\begin{center}
\includegraphics[width=\figwidth,clip]{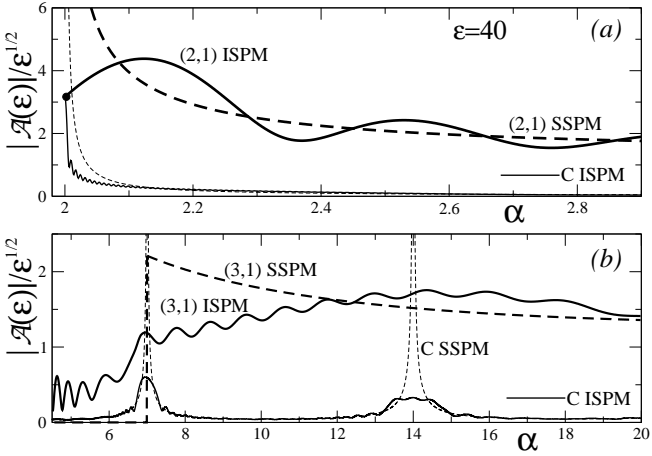}
\end{center}
\caption{\label{fig2} 
Moduli of the scaled ISPM (solid) and SSPM
(dashed curve) amplitudes 
$|\mathcal{A}_{\rm PO}(\varepsilon)|$ as functions of $\alpha$ 
for the primitive ($M=1$)
circle $C$ [Eq.~(\ref{ampra0})], diameter $(2,1)$ and
triangle-like $(3,1)$ [Eq.~(\ref{ampra1})] POs, in units of
$\varepsilon^{1/2}$ at the scaled energy $\varepsilon=40$.  The 
panels show {\it (a)} the HO limits ($\alpha\to 2$) 
for the
circle $C$ (thin) and diameter $(2,1)$ (thick curve) orbits (the
filled circle denotes one half of the HO amplitude (\ref{holimC}) at
$\alpha=2$); and {\it (b)}  
the circle $C$ (thin) and triangle-like
$(3,1)$ (thick curve) POs.} 
\end{figure}

A remarkable enhancement of the ISPM amplitudes in PO sum for
the oscillating level density (\ref{dgrake}) and
shell correction energy (\ref{escscl}) due to the bifurcation
(symmetry breaking)
is typically expected for some short periodic orbits.
In Fig.~\ref{fig2}, the scaled
amplitudes $|\mathcal{A}_{\rm PO}|$, divided by $\varepsilon^{1/2}$ to
normalize the energy dependence for $\mathcal{K}=1$ orbits, are
presented for several shortest POs as functions of the power
parameter $\alpha$ in order to show the typical
bifurcation enhancement phenomena. 
In Fig.~\ref{fig2}{\it (a)}, the enhancement of the
primitive diameter $(2,1)$ amplitudes $|\mathcal{A}_{(2,1)}|$ 
[Eq.~(\ref{ampra1})], and those $|\mathcal{A}_{MC}|$
[Eq.~(\ref{ampra0})] for the primitive circle orbit $C$ are clearly
seen in the HO limit $\alpha \rightarrow 2$; see also Eq.~(\ref{dgrakeps}).
Figure~\ref{fig2}{\it (b)} shows the enhancement of the
shortest orbit $C$ around the bifurcation point $\alpha=7$, and the
birth of the triangle-like orbit $(3,1)$ there.  Note that the ISPM
amplitude $|\mathcal{A}_{(3,1)}|$ [Eq.~(\ref{ampra1})] of the $(3,1)$
orbit keeps its magnitude up to rather a large value of $\alpha$ above
the bifurcation ($\alpha > \alpha_{\rm bif}$).  The ISPM amplitude for
the circle orbit $C$ exhibits a remarkable enhancement at the
bifurcation point $\alpha=7$.  The divergence of its SSPM amplitude
at the bifurcation point is successfully removed.  As also seen from
Fig.~\ref{fig2}{\it (b)}, the ISPM amplitude for the $(3,1)$ PO
is continuously changed through this bifurcation, in contrast to the
discontinuity of the SSPM amplitude.  This orbit exists, in fact, only
at $\alpha \geq 7$, and the amplitude in the region $\alpha<7$ is due
to the formal stationary point which has no direct sense in the
classical dynamics.  Therefore, the corresponding PO is called usually
as a ghost orbit
\cite{book,spheroid,maf}.  An oscillatory behavior of the amplitude
$|\mathcal{A}_{(3,1)}|$ in the ghost region far from the bifurcation
has no physical significance, since it is washed out in the 
Gaussian-averaged
level density by an rapidly oscillating phase of the
complex amplitude $\mathcal{A}_{(3,1)}$ 
\cite{spheroid,maf}.  These ghost amplitude oscillations are
suppressed even more by using higher order expansions in the phase and
amplitudes in a more precise ISPM 
\cite{spheroid}.

\begin{figure}
\begin{center}
\includegraphics[width=\figwidth,clip]{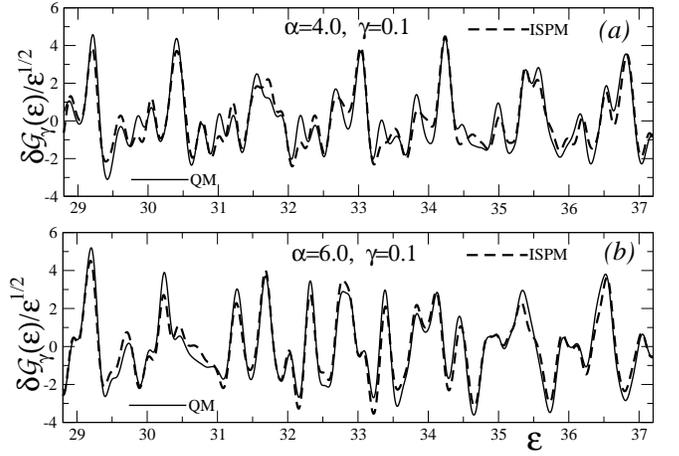}
\end{center}
\caption{\label{fig3} 
The oscillating part of the level density
$\delta\mathcal{G}_\gamma(\varepsilon)$ in units of
$\varepsilon^{1/2}$ vs the
scaled energy $\varepsilon$ for $\alpha=4.0$ {\it (a)},
and $\alpha=6.0$ {\it (b)}, at the (dimensionless) 
width parameter 
$\gamma=0.1$ in 
the Gaussian averaging over the scaled energies; 
the solid and dashed curves are the quantum-mechanical 
and semiclassical ISPM results, respectively.
}
\end{figure}

\begin{figure}
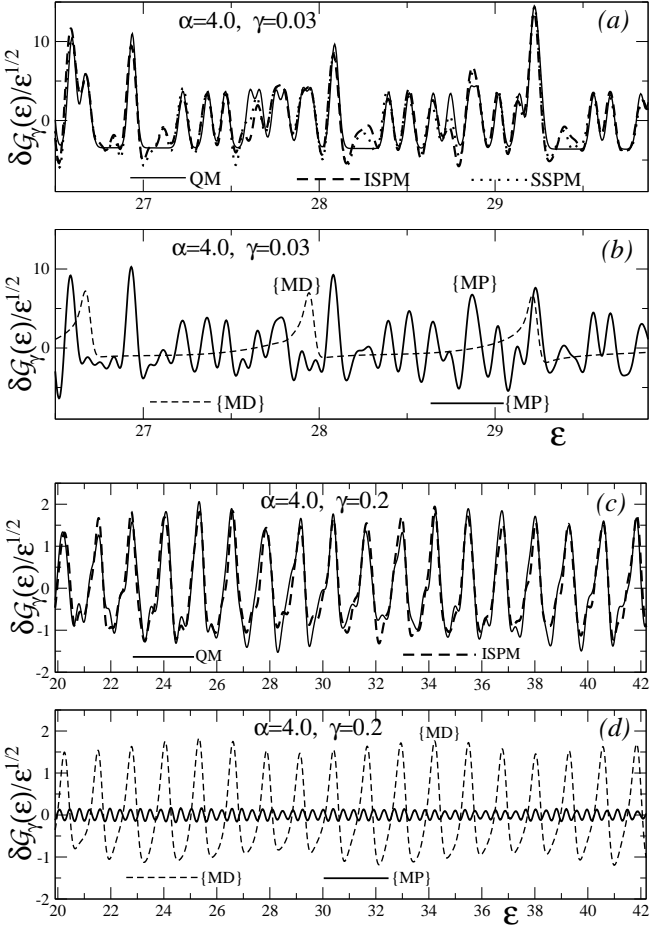

\begin{center}
\includegraphics[width=\figwidth,clip]{fig4ab.eps}\\[12pt]
\includegraphics[width=\figwidth,clip]{fig4cd.eps}
\end{center}
\caption{\label{fig4} 
The same as in Fig.~\ref{fig3} for $\alpha=4.0$ but with 
other width parameters, $\gamma=0.03$
{\it (a,b)} and $\gamma=0.2$  
{\it (c,d)}.
{\it Panels (a,c)}: 
the solid, dashed and dotted lines are the QM,
ISPM and SSPM [ 
the panel {\it (a)}] results. 
{\it Panels (b,d)}: $\{MD\}$ (dashed) is the contribution of the diameters
(including their repetitions) and $\{MP\}$ (thin solid) for other
$\mathcal{K}=1$ polygon-like POs.
}
\end{figure}

\begin{figure}
\begin{center}
\includegraphics[width=\figwidth,clip]{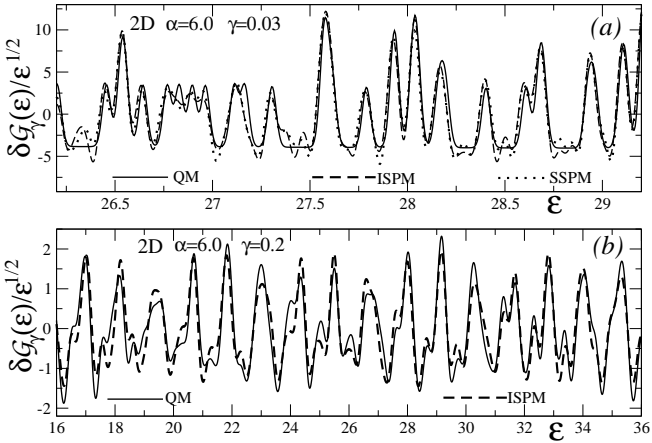}
\end{center}
\caption{\label{fig5} 
The same as in Fig.~\ref{fig3} for $\alpha=6.0$, but with 
other width parameters, $\gamma=0.03$ 
{\it (a)}, and $\gamma=0.2$  {\it (b)}. 
}
\end{figure}

\begin{figure}
\begin{center}
\includegraphics[width=\figwidth,clip]{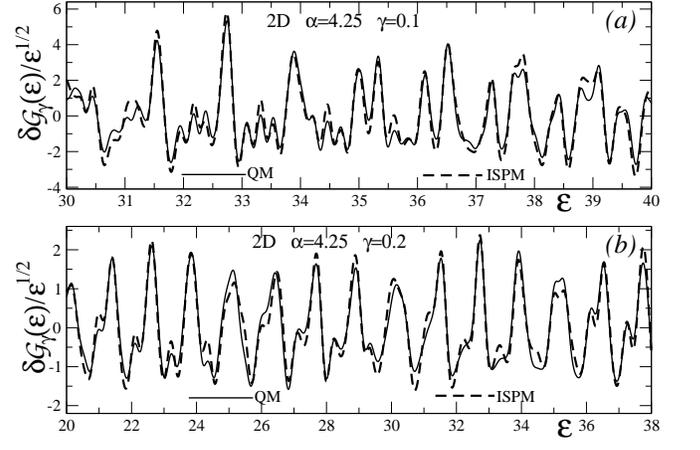}
\end{center}
\caption{\label{fig6} 
The same as in Fig.~\ref{fig5} for $\alpha=4.25$, but with 
the width parameters, $\gamma=0.1$ {\it (a)}, and $\gamma=0.2$ {\it (b)}.
}
\end{figure}

\begin{figure}
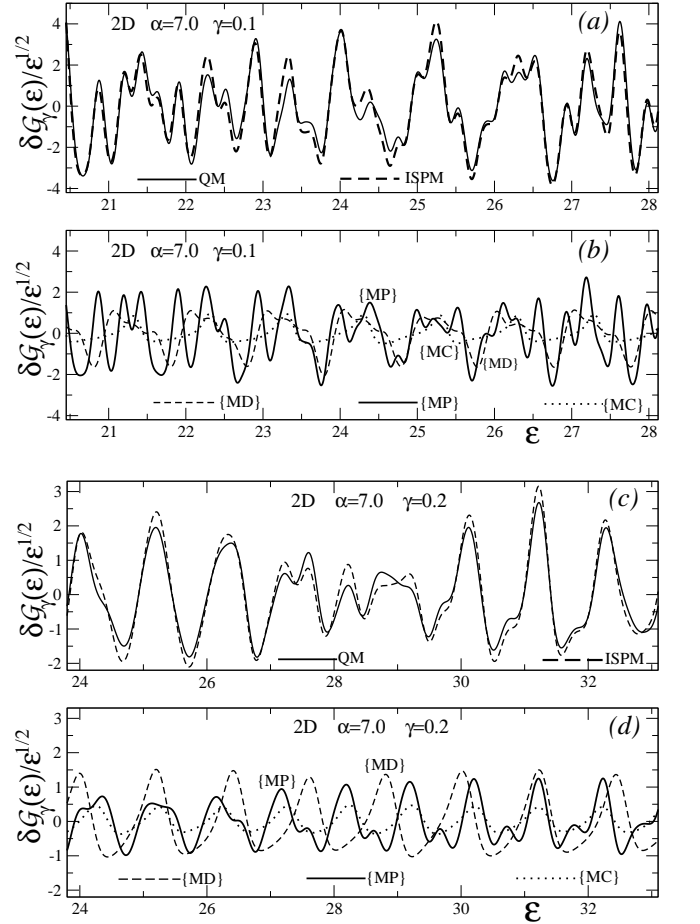

\begin{center}
\includegraphics[width=\figwidth,clip]{fig7ab.eps} \\[12pt]
\includegraphics[width=\figwidth,clip]{fig7cd.eps}
\end{center}
\caption{\label{fig7} 
The same as in Fig.~\ref{fig4}, but for $\alpha=7.0$ and 
other width parameters, $\gamma=0.1$ 
{\it (a,b)} and $\gamma=0.2$ {\it (c,d)}. 
}
\end{figure}

Figures~\ref{fig3}--\ref{fig7} show the oscillating part of
the semiclassical scaled-energy level density
$\delta\mathcal{G}_\gamma(\varepsilon)$
[Eq.~(\ref{avdenra})] in units of $\varepsilon^{1/2}$ as functions of the
scaled energy $\varepsilon$ for several values of the
power parameter $\alpha$ and the Gaussian width $\gamma$.
The ISPM semiclassical results show good agreement with the quantum
mechanical (QM) ones  
for a transition from the gross to fine resolutions 
of the spectra.
The QM calculations are carried out by the use of the standard
Strutinsky averaging over the scaled energy $\varepsilon$, in which
we find a good plateau around the Gaussian averaging width
${\widetilde \gamma}=2-3$ with the even curvature correction
polynomials of 4th to 8th powers.

For the powers $\alpha=4.0$ and $6.0$, one finds a good agreement with
the SSPM asymptotic behavior [Figs.~\ref{fig4}{\it (a)} and
\ref{fig5}{\it (a)}]
because they are sufficiently far from the
bifurcation points $\alpha=4.25$ and $7.0$ which correspond to the birth
of the star-like $(5,2)$ and triangle-like $(3,1)$ POs
(Figs.~\ref{fig6} and \ref{fig7}).
For the gross shell structure ($\gamma \approx 0.2$ at $\alpha=4.0$),
only the shortest orbits (mainly a few shortest diameters) give
the leading contributions.
(This is in contrast to the 3D case where the
circular orbits become also important \cite{trap,book}.)
For instance, the gross shell structure in terms of the shortest POs
for $\alpha=6.0-7.0$ manifests at larger $\gamma \gtrsim 0.3$, unlike for
the powers $\alpha=4.0-4.25$.  With decreasing $\gamma$ and
increasing $\alpha$, the POs
for larger scaled periods $\tau$ [or actions $S$, see
Eq.~(\ref{actionsc})] become more significant [cf.
Figs.~\ref{fig4}{\it (b,d)} and \ref{fig7}{\it (b,d)}].
In the
case of the fine shell structure (e.g., $\gamma \approx 0.03$) the
dominant contributions are due to the bifurcating
$\mathcal{K}=1$ POs [polygon-like POs denoted by $\{MP\}$; see 
Fig.~\ref{fig4}{\it (b)}].
(This is similar to the situations in the elliptic
\cite{ellipse} and spheroidal \cite{spheroid} cavities, and in the IHH 
potential \cite{maf}.)  However, the interference of these much 
longer one-parametric POs
[such as $M(7,3)$ for $\alpha=4.0$ or $M(5,2)$ for $\alpha=6.0$]
with a lot of the $M(2,1)$ diameters explain some peaks, too. 
For smaller $\alpha=4.0$ and $4.25$, the
circle orbit contributions are not shown
because they are insignificant at these power parameters 
in the 2D case.
(This is different situation from the 3D case, see Ref.~\cite{trap} for
the trace formulas based on the uniform approximation using the
classical perturbation approach \cite{creagh96,book}.)  
These contributions into the trace
formula (\ref{avdenra}) are increasing functions of $\alpha$, and they
become significant at $\alpha \gtrsim 7$ even for the 2D case
[Fig.~\ref{fig7}{\it (b,d)}].  An intermediate situation between the
gross- and fine- shell structures where all of POs become significant
are shown too at $\gamma=0.1$ in Figs. \ref{fig3} and \ref{fig6}, and
at $\gamma=0.1$
and $0.2$ in Fig.~\ref{fig7}{\it (b,d)}.
Our full analytical
expressions (accessible for any long periodic orbits) for the classical 
PO characteristics
at $\alpha = 4$ and $6$  are quite useful in the
simple ISPM calculations of the oscillating level density 
with a good accuracy up to the
fine spectrum-structure resolutions by using, for instance,
$\gamma\approx 0.03$ and $0.1$.
Figures~\ref{fig6} and
\ref{fig7} show a nice agreement of the fine-resolved semiclassical
and quantum level densities
$\delta\mathcal{G}_\gamma(\varepsilon)$ as functions of the scaled
energy $\varepsilon$ at the critical bifurcation points $\alpha=4.25$
and $7.0$ for the births of the star-like $(5,2)$ and 
triangle-like $(3,1)$ orbits, respectively.

\begin{figure}
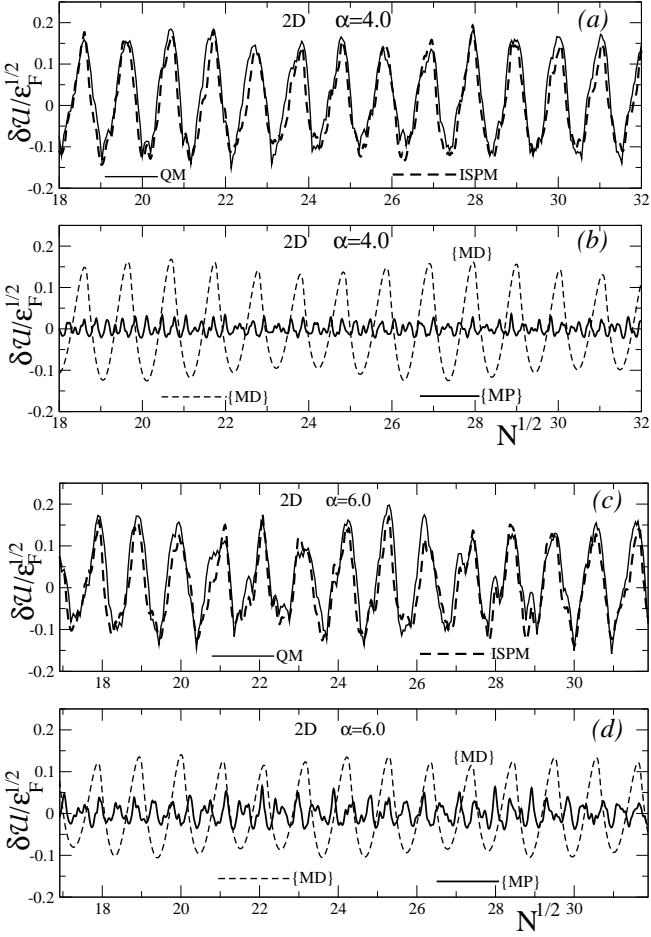

\begin{center}
\includegraphics[width=\figwidth,clip]{fig8ab.eps} 
\\[12pt]
\includegraphics[width=\figwidth,clip]{fig8cd.eps}
\end{center}
\caption{\label{fig8}
Scaled shell correction energies $\delta\mathcal{U}$,
normalized by the factor $\varepsilon_F^{-1/2}$, as functions of
square root of the particle number
$N^{1/2}$ at the values of $\alpha$, where the full analytical
formulas are obtained for $\alpha=4.0$ {\it (a,b)} and 
$\alpha=6.0$ {\it (c,d)}.
{\it Panels (a,c)}: QM (solid curve) represents the
quantum-mechanical results using the Strutinsky SCM, and ISPM
(dashed curve) shows the semiclassical result using the TF
approximation in the calculation of $N(\varepsilon_F)$ by
Eqs.~(\ref{partnum2}) and (\ref{scldenstyra}).
{\it Panels (b,d)}: the contributions of several POs into
the shell correction energy $\delta\mathcal{U}$ are
shown.  Other
notations are the same as in Figs.~\ref{fig4} and \ref{fig7}.  }
\end{figure}

\begin{figure}
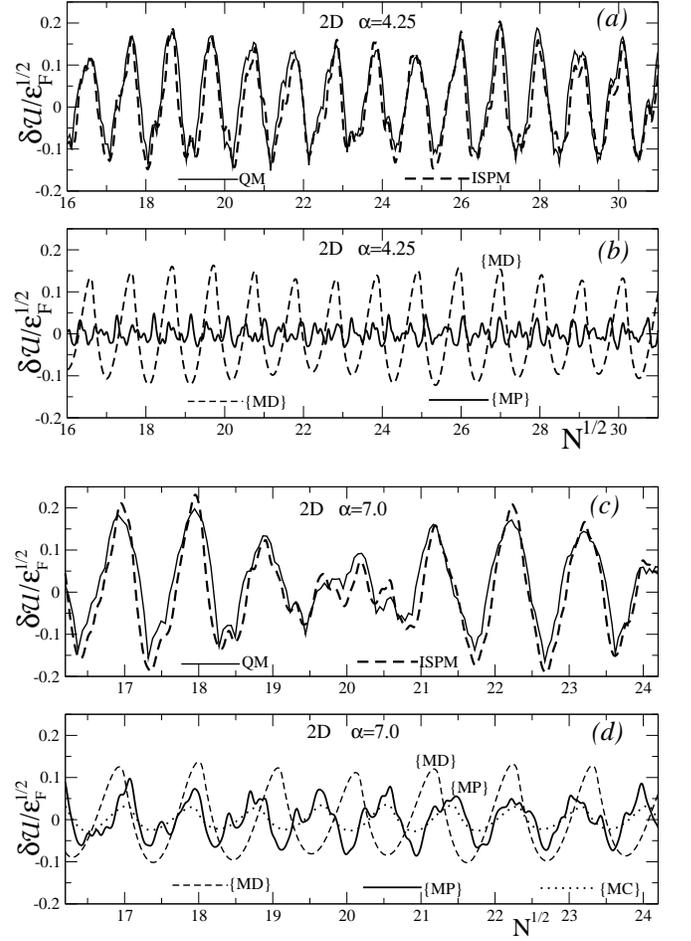

\begin{center}
\includegraphics[width=\figwidth,clip]{fig9ab.eps}
\\[12pt]
\includegraphics[width=\figwidth,clip]{fig9cd.eps}
\end{center}
\caption{\label{fig9}
The same as in Fig.~\ref{fig8}; $\alpha=4.25$ for the (5,2) bifurcation, 
and $\alpha=7.0$ for the (3,1) bifurcation are shown in panels {\it (a,b)}
and {\it (c,d)}, respectively.}
\end{figure}

Figures~\ref{fig8} and \ref{fig9}  
show the scaled shell correction energies $\delta\mathcal{U}$
[Eqs.~(\ref{escscl2}) for the semiclassical
 and (\ref{qmrelscm2}) for the quantum results], normalized
by the factor $\varepsilon_F^{-1/2}$, 
as functions of the particle number variable $N^{1/2}$.
A good plateau is realized for the QM calculations of the
scaled shell-correction energies
[see the first equation in Eq.~(\ref{qmrelscm2})]
near the same averaging parameters ${\widetilde \gamma}$ and
curvature corrections as mentioned above. 
In the semiclassical calculations, the Fermi level 
$\varepsilon_{{}_{\!F}}$ is
determined by the particle number conservation (\ref{partnum2}) 
with using the coarse-grained scaled-energy POT level density, 
\begin{equation}\label{scldenstyra}
\mathcal{G}_{\gamma,\;{\rm scl}}(\varepsilon)
=\mathcal{G}_{\rm TF}(\varepsilon) +
\sum_{\mathcal{K}=0}^1
\delta\mathcal{G}^{(\mathcal{K})}_{\gamma,\;scl}(\varepsilon).
\end{equation}
The oscillating ISPM components $\delta
\mathcal{G}^{(\mathcal{K})}_{\gamma,\;scl}(\varepsilon)$ 
are given
by Eqs.~(\ref{avdenra}) and (\ref{dgrakeps}).
We evaluated the Fermi level $\varepsilon_{{}_{\!F}}(N)$ by varying the
averaging width $\gamma$ and found that
there is no essential sensitivity 
within the interval of smaller $\gamma$ ($\gamma \approx 0.1-0.2$).
Moreover, even the TF density
$\mathcal{G}_{\rm TF}(\varepsilon)$ [Eq.~(\ref{smoothef}] 
in Eq.~(\ref{partnum2}) with $\mathcal{G}(\varepsilon)\approx
\mathcal{G}_{\rm TF}(\varepsilon)$ provides us a
good value of $\varepsilon_{{}_{\!F}}$
in the POT calculations of the shell correction energies 
(\ref{escscl2}).
The PO sums at $\alpha=7.0$ 
converge for the shell correction density (\ref{avdenra})
by using the averaging width $\gamma=0.2$ of a fine shell-structure
resolution, and for the shell correction energies (\ref{escscl2})
with taking into account the same major simplest POs [about 4 repetition
numbers ($M=4$) for the circle and diameter orbits, and 
a few first simplest
other $\mathcal{K}=1$ ($P$) POs, such as $(3,1)$, $(5,2)$, $(7,3)$ and $(8,3)$;
cf. Figs. \ref{fig9}(c,d) with \ref{fig7}(c,d)].  
For smaller diffuseness, $4 \lesssim \alpha \lesssim 6$, one has a 
similar PO convergence relation with the same 
$\gamma \approx 0.2$, but with much smaller contributions of the 
circular orbits. However, the dominating ($\mathcal{K}=1$) PO families
(P) are the $(5,2), (7,3)$ and $(7,3)$ POs at $\alpha=4.25-6.0$ and $4.0$,
respectively [Figs. \ref{fig8}(a,b) and \ref{fig4}(c,d)].
As seen from Figs.~\ref{fig6},
\ref{fig7} and \ref{fig9}, we obtain a nice agreement between the
semiclassical (ISPM, dashed) and quantum (QM, solid curve) 
results exactly
at the bifurcations $\alpha=4.25$ and $7.0$.  Notice that the dominating
contributions in these semiclassical results at the bifurcation point
$\alpha=7.0$ are coming from the interference of the bifurcating circle
$C$ and newborn $(3,1)$ orbits with the simplest diameters.  As shown
typically in Figs.~\ref{fig7}{\it (d)} and \ref{fig9}{\it (d)},
one can see that the circle $C$ and triangle-like
$(3,1)$ orbits are mainly in phase, but the diameter $(2,1)$ is sometimes in
phase to them and sometimes out of phase.  Thus, the occurrence
of a characteristic beating pattern in the level density amplitude
at $\alpha=7.0$ is due to the interference of
the bifurcating orbits $C$ and $(3,1)$ with the shortest diameter $(2,1)$
having all the amplitude of the same order in magnitude but different
phases.  The bifurcating circle $2C$ and
star-like $(5,2)$ orbits [as expected from the enhancement of the
amplitudes of the circular $C$ and triangular-like
$(3,1)$ POs in Fig.~\ref{fig2}] are more important for
$\alpha=4.25$, though the primitive diameters become significant much
compared to the bifurcation case $\alpha=7.0$.  The POs $(3,1)$ and
$(5,2)$ yield more contributions near their bifurcation values of
$\alpha$, and even more on the right-hand side
($\alpha \gtrsim \alpha_{\rm
bif}$) in a wide region of $\alpha$ as
mentioned above.  The bifurcation parent-daughter partner orbits
$\{C$, $(3,1)\}$ and
$\{2C$, $(5,2)\}$, taken together with the simple 
diameter $(2,1)$, give
essential ISPM contributions of about the same order of magnitude
in Figs.~\ref{fig6}, \ref{fig7} and \ref{fig9}; as seen for
example in 
Figs.~\ref{fig7}{\it (b,d)} and \ref{fig9}{\it (d)}
for the same $\alpha=7.0$.
The diameter ISPM contributions are close to the SSPM asymptotic ones
near the bifurcation points $\alpha=7.0$ and $4.25$ (as for 
$\alpha=4.0$ and $6.0$) because
they are sufficiently far from their single
symmetry-breaking point at 
the harmonic oscillator value $\alpha=2$.
\begin{figure}
\begin{center}
\includegraphics[width=\figwidth]{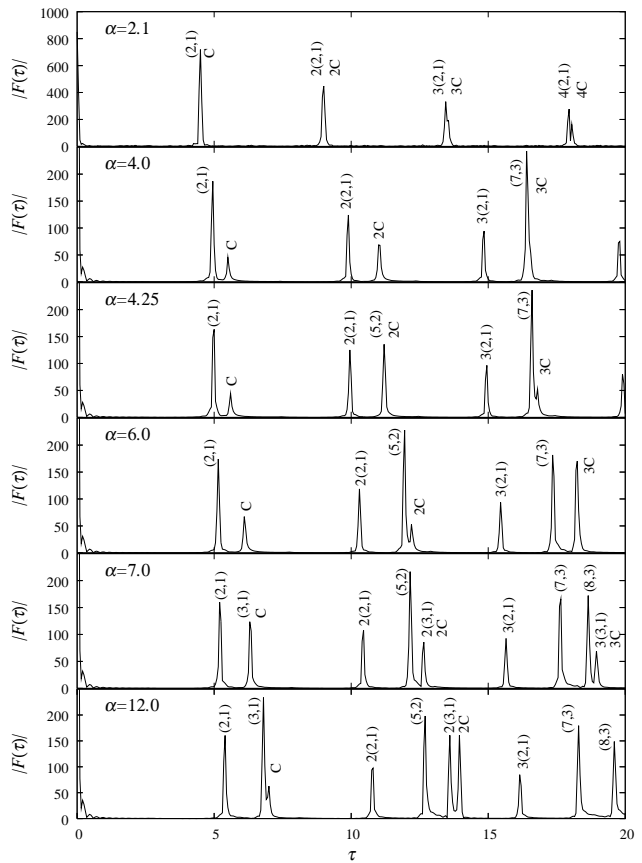}
\end{center}
\caption{\label{fig10}
Moduli of the Fourier transform $|F(\tau)|$ of 
the quantum scaled-energy level density (\ref{fourier_power_qm})
as functions of the dimensionless variable $\tau$  
are plotted for several values of $\alpha$;
$MC$ and $M(n_r,n_\varphi)$ indicate the classical POs
corresponding to each peak (see Fig.\ \ref{fig1}).
}
\end{figure}

Figure~\ref{fig10} shows the Fourier transform of the
quantum-mechanical scaled-energy level density
[Eq.~(\ref{fourier_power_qm})].  For a smaller $\alpha=2.1$, the
diameter $(2,1)$ orbit gives the dominant contribution to the
gross-shell structure as the shortest POs; see the peak at $\tau\sim
5.0$.  With increasing $\alpha$, the amplitude of the circle orbit
becomes again larger due to a prominent enhancement around the
bifurcation point ($\tau\sim 6.2$ at $\alpha_{\rm bif}=7.0$).  Notice
that the newborn POs $(3,1)$, $(5,2)$, $(7,3)$ and $(8,3)$ give
comparable contributions at $\alpha=7.0$ [similarly, $(5,2)$ and
$(7,3)$ for the bifurcation $\alpha=4.25$] in nice agreement with the
quantum Fourier spectra in Fig.~\ref{fig10}.  The contributions of the
newborn triangle-like orbit family $(3,1)$ having  
relatively a smaller scaled period $\tau_{{}_{\!(3,1)}}$ and
higher degeneracy $\mathcal{K}=1$ become important and dominating for larger
$\alpha \gtrsim \alpha_{\rm bif}=7$.  The newborn $(3,1)$ peak cannot be
distinguished from the parent circle $C$ orbit near the bifurcation
point $\alpha_{\rm bif}$ as well as the diameter and circle orbits at
$\alpha$ close to the HO limit, see Ref.\ \cite{migdal}.  We emphasize
that the shell correction energies $\delta \mathcal{U}$ are similar to the
oscillating parts of the level 
densities coarse-grained over the spectrum by
using the Gaussian width $\gamma=0.2$ at $\alpha=4.0$ and $4.25$, which
have mainly the gross-shell structure due to the shortest diameters.
However, for this $\gamma$, the fine-resolved shell
structures (due to
their interference with the other polygon-like and circular POs) are
pronounced at larger powers near $\alpha=6.0$, and especially, $7.0$.

Notice also that we do not show the numerical comparison of the ISPM 
[Eqs.\ (\ref{avdenra}) and (\ref{escscl2})] vs quantum (Ref.\ \cite{book}) 
results for the HO limit $\alpha \to 2$ because they are 
exactly coincide, as shown analytically around Eq.~(\ref{holimC}).

\section{Conclusions}
\label{sec5}

We presented a semiclassical theory of quantum oscillations of the
level density and energy shell corrections for a class of radial
power-law potentials which turn out as good approximations to the
realistic Woods-Saxon potential in the spatial region where 
the particles are
bound.  The advantage of the RPL potentials is that, in spite of its
diffuse surface, the classical dynamics scaling
with simple powers of
the energy simplifies greatly the analytical
POT calculations. The quantum Fourier
spectra yield directly the contributions of the leading classical POs
with the specific periods and actions into the trace formulas.

We described the main PO properties of the classical dynamics
in the RPL potentials as the key quantities of the POT.  Taking the
simplest two-dimensional RPL Hamiltonian we developed 
the semiclassical trace formulae
for any its power $\alpha$, and studied various limits of $\alpha$ (the
harmonic oscillator potential for $\alpha=2$ and the cavity potential
for $\alpha\to\infty$).  The completely analytical results were
obtained for the RPL powers $\alpha=4$ and $6$.  This can be applied
for both 2D and 3D cases and allow us to far-going fine-resolved shell
structures at $\gamma=0.03-0.1$.  This POT is based upon extended
Gutzwiller's trace formula, that connects the level density of a
quantum system to a sum over POs of the corresponding classical
system.  It was applied to express the shell correction energy $\delta
U$ of a finite fermion system in terms of POs.  We obtained good
agreement between the ISPM semiclassical and quantum-mechanical
results for the level densities and energy shell corrections at
several critical powers of the RPL potentials.  For the powers
$\alpha=4$ and $6$, we found also good agreement of the ISPM trace
formulas with the SSPM ones.  The strong amplitude-enhancement
phenomena at the bifurcation points $\alpha=7$ and $4.25$ in the
oscillating (shell) components of the level density and energy were
observed in the remarkable agreement with the peaks of the Fourier
spectra.  We found a significant influence of the PO bifurcations on
the main characteristics (oscillating components of the level
densities and energy-shell corrections) of a fermionic quantum
system.  They leave signatures in its energy spectrum (visualized,
e.g., by its Fourier transform), and hence, its shell structure.  We
have presented a general method to incorporate bifurcations in the
POT, employing the ISPM based on the catastrophe theory of Fedoryuk
and Maslov, and hereby, overcoming the divergence of the semiclassical
amplitudes of the Gutzwiller theory and their discontinuity in the
Berry\&Tabor approach at bifurcations.  The improved semiclassical
amplitudes typically exhibit a clear enhancement near a bifurcation
and on right side of it, where new orbits emerge, which is of the order
$\hbar^{-1/2}$ in the semiclassical parameter $\hbar$.  This, in turn,
leads to the enhanced shell structure effects.  Bifurcations are
treated, again, in the ISPM leading to the semiclassical enhancement
of the orbit amplitudes.  The trace formulae are presented 
numerically to
show good agreement with the quantum-mechanical level density
oscillations for the gross- (coarse-grained
with larger averaging width
$\gamma$ and a few shortest POs), and the fine-resolved (with
smaller $\gamma$
and longer bifurcating POs) shell structures.
The PO structure of the shell-correction energies 
is similar to that of the coarse-grained 
densities for smaller powers $\alpha=4$ --
$4.25$, and 
of the fine-resolved densities for larger $\alpha \gtrsim 6$
 at the same $\gamma\approx 0.2$. The
fine-resolved and coarse-grained  
shell structures were found at the same $\alpha $ 
in the corresponding averaged oscillating 
densities at smaller
width parameters $\gamma=0.03$ -- $0.1$  
and  at larger ones $\gamma \gtrsim 0.2$ -- $0.3$, respectively. 
The fine-resolved shell structure for larger powers, 
$\alpha \gtrsim 6$, 
occurs in a larger interval, $\gamma=0.003$ -- $ 0.2$, including the essential
contributions of the circle orbits along with the polygon-like and diameter
orbits.  Full explicit
analytical expressions for the diameters and circle orbit
contributions into the trace formula as functions of the diffuseness
potential parameter $\alpha$ are specified too.

For prospectives, we intend a further study of shell structures in the
3D RPL potentials, within the ISPM and uniform approximations
to treat the bifurcations, by varying continuously the power parameter
$\alpha$ from 2 (harmonic oscillator) to $\infty$ (spherical
billiards).

\acknowledgments

Authors thank Profs. M. Brack and K. Matsuyanagi for many valuable  
discussions.

\appendix
\section{The stability factor,  bifurcation powers and frequencies}
\label{appA}
 
Let us consider in more details the non-linear classical dynamics in
the RPL Hamiltonian (\ref{ramodra}) for any real $\alpha\geq 2$.  The
critical values of the radial coordinate $r=r_{{}_{\!C}}$ and angular
momentum $L=L_C$ for the circle orbit ($C$) are determined by the
solutions of the system of the two equations with respect to $r$ and
$L\;$:
\begin{equation}
 \mathcal{F}(r,L)=0,\quad \frac{\partial\mathcal{F}}{\partial r}=0,
\quad \text{where} \quad \mathcal{F}(r,L) \equiv p_r^2(r,L)\;,
\label{eqrc}
\end{equation}
see Eq.~(\ref{pr}).  In the internal region where the stable orbits in
the radial direction exist, one has a nonzero
$\mathcal{F}_C^{\prime\prime}=
\partial^2  \mathcal{F}\left(r_{{}_{\!C}},L_C\right)/\partial r^2
< 0$. First equation in
Eq.~(\ref{eqrc}) means that there is no radial velocity, $\dot{r}=0$,
and the next equation is that the radial force is equilibrating by the
centrifugal force.  For the Hamiltonian (\ref{ramodra}), the
solutions of the two these equations are the radius $r_{{}_{\!C}}$ and
angular momentum $L_C$ \cite{aritapap},
\begin{equation}
r_{{}_{\!C}}=R_0\left(\frac{2E}{(2+\alpha)E_0}\right)^{1/\alpha},\qquad 
L_C=p(r_{{}_{\!C}})r_{{}_{\!C}}\;. 
\label{rcLcd2Fcra}
\end{equation}
Using Eq.~(\ref{freq}) at $L=L_C$ for the rotational frequency,
$\omega_{{}_{\!C}}=\omega_\varphi(L=L_C)=L_C/(m r_{{}_{\! C}}^2)$,  
and  (\ref{rcLcd2Fcra}) for
$r_{{}_{\!C}}$ and $L_C$, 
one finds \cite{aritapap}
\begin{equation}
\omega_{{}_{\!C}}=\sqrt{\frac{\alpha E_0}{mR_0^2}}\,
\left(\frac{2E}{(2+\alpha)E_0}\right)^{1/2-1/\alpha}\;.
\label{omtcra}
\end{equation} 
Applying now the second order expansion in $r-r_{{}_{\!C}}$ to Eq.~(\ref{pr}),
one gets the first-order ordinary differential equation
for the radial CT $r(t)$ locally near the circle PO $r=r_{{}_{\!C}}$:
\begin{equation}
\dot{r}=\pm \sqrt{\frac{\mathcal{F}_C^{\prime\prime}}{2m^2}}\,
\left(r-r_{{}_{\!C}}\right)\;.
\label{rtloceq}
\end{equation}
Integrating the dynamical equation in Eq.~(\ref{rtloceq}), one obtains
\begin{equation}
r(t)=r_{{}_{\!C}}+\left(r^\prime-r_{{}_{\!C}}\right)\,\, 
\exp\left(\pm\sqrt{\frac{\mathcal{F}_C^{\prime\prime}}{2\,m^2}}\,\,t
\right)\;,
\label{rtlocsol}
\end{equation}
where $r^\prime=r\left(t=t^\prime=0\right)$.
In the stable case, $\mathcal{F}_C^{\prime\prime}<0$ in 
Eq.~(\ref{rtlocsol}) for the CT $r(t)$ locally near the circle
orbit $r=r_{{}_{\!C}}$, one writes
\begin{equation}
r(t)=r_{{}_{\!C}}+\left(r'-r_{{}_{\!C}}\right)\,
\exp\left(\pm i \varOmega_C\,t\right)\;,
\label{rtlocsol1}
\end{equation}
where $\varOmega_C$ is a positive radial frequency $\omega_r$ at $L=L_C$
[Eq.~(\ref{freq})],
\begin{equation}
\varOmega_C=\sqrt{\left|\mathcal{F}_C^{\prime\prime}/(2\,m^2)\,\right|}=
\omega_r(L=L_C)\;.
\label{omc}
\end{equation}
For the Hamiltonian (\ref{ramodra}), this quantity is given by 
\cite{aritapap}
\begin{equation}
\varOmega_C=\sqrt{\frac{2\alpha E}{mR_0^2}}\,
\left[\frac{(2+\alpha)E_0}{2E}\right]^{1/\alpha}>0\;.
\label{Omrcra}
\end{equation}
{}From Eq.~(\ref{rtlocsol1}) after the period $T_{C}$
along the primitive circle orbit,
\begin{equation}
T_{C}=t^{\prime\prime}-t^\prime=t^{\prime\prime}=
\frac{2 \pi}{\omega_{{}_{\!C}}}\;, 
\label{periodTc}
\end{equation}
one finds
\begin{equation}
\delta r^{\prime\prime}\equiv r^{\prime\prime}-r_{{}_{\!C}}
= \delta r^\prime\, \exp\left(\pm i
  \varOmega_C\,T_{C}\right),
\quad \delta r^\prime=r^\prime-r_{{}_{\!C}}\;.
\label{dr}
\end{equation}
The eigenvalues of the stability matrix ${\cal M}_{C}$ 
for $M=1$ in Eq.~(\ref{fgutzC}) are given by \cite{book}
\begin{equation}
\left(\frac{\partial r^{\prime\prime}}{\partial r^\prime}\right)_{p_r^\prime}
\!\!=
\exp{\left(i  \varOmega_C\,T_{C}\right)},\quad
\left(\frac{\partial p_r^{\prime\prime}}{\partial p_r^\prime}\right)_{r^\prime}
\!\!=
\exp{\left(- i  \varOmega_C\,T_{C}\right)}\;.
\label{M11}
\end{equation}
These two eigenvalues of the stability matrix are complex conjugated 
in agreement with its general properties. 
As $\varOmega_C$ is real [$\varOmega_C>0$, according to 
Eqs.~(\ref{omc}) and (\ref{Omrcra})]
the circle orbit is isolated stable PO. 
Substituting the expressions (\ref{M11}) into the first equation in
Eq.~(\ref{fgutzC}) and using 
Eqs.~(\ref{periodTc}) for the period $T_C$, (\ref{omtcra}) and 
(\ref{Omrcra}) for the C orbit frequencies $\omega_{{}_{\!C}}$ and $\varOmega_C$,
relatively,
one obtains the last equation in Eq.~(\ref{fgutzC}) 
for the stability factor $F_{MC}$.

\section{Scaling properties}
\label{appB}

For convenience, let us consider the classical dynamics in terms of the
variables in dimensionless units $m=R_0=E_0=1$.
Due to the scaling property (\ref{scaling}) for the classical dynamics in
the Hamiltonian (\ref{ramodra}),
the energy dependence
of the action $I_r(\varepsilon)$ [Eq.~(\ref{actionvar})],
the angular momentum $L(\varepsilon)$, the frequency $\omega_r(\varepsilon)$ 
[Eq.~(\ref{freq})] and
the curvature $K(\varepsilon)$ [Eq.~(\ref{curv})]
can be expressed in terms of the simple powers of the 
scaled energy $\varepsilon$,
\begin{equation}
\varepsilon=E^{1/\alpha+1/2}\;.
\label{scalE}
\end{equation}
In particular, one can express these classical quantities
through their values at $\varepsilon=1$ ($E=1$),
\begin{gather}
I_i=I_i(1)\varepsilon,\quad
L=L(1)\varepsilon,\quad
\omega_r^{-1}=\omega_r^{-1}(1)\varepsilon^{(2-\alpha)/(2+\alpha)},
\nonumber\\
K=K(1)/\varepsilon\;.
\label{scaliolk}
\end{gather}
Therefore, due to
the scaling properties (\ref{scaling}) and (\ref{scaliolk}), we need
to calculate these classical dynamical quantities only at one
value of the energy $\varepsilon=1$.  For simplicity of notations,
we shall omit the argument
$\varepsilon=1$ everywhere, if it is not lead to misunderstandings.

The radial action $I_r(L,E)$ [Eq.~(\ref{actionvar})] 
can be expressed explicitly in terms of the frequencies $\omega_\varphi$
and $\omega_r$ [Eq.~(\ref{freq})], and their ratio $f(L)$
[Eq.~(\ref{fldef})],
\begin{equation}
I_r = \frac{2 \alpha}{\alpha+2}\,\omega_r^{-1} - L\,f(L)\;. 
\label{actra}
\end{equation}
To prove this identity, we express  Eq.~(\ref{freq}) 
for $\omega_r^{-1}$ in terms of the determinant,
\begin{equation}
\omega_r^{-1}=
\frac{\partial(I_r,L)}{\partial(E,L)}=\frac{\partial I_r}{\partial E}-
\frac{\partial I_r}{\partial L}\;\frac{\partial L}{\partial E}\;.
\label{freq1}
\end{equation}
Calculating directly the derivatives in this equation by 
using Eq.~(\ref{scaliolk}),
one obtains the expression for $\omega_r^{-1}(1)$.  Solving then
this equation with respect to $I_r(1)$, one arrives at Eq.~(\ref{actra}).
Differentiating the identity (\ref{actra}) term by term 
over $L$ and using the definition for the ratio
of frequencies $f(L)$ [Eq.~(\ref{fldef})], for the 
curvature (\ref{curv}) one finally obtains 
\begin{align}
K&=-\frac{2\alpha}{(\alpha+2) L}\;\frac{\partial \omega_r^{-1}}{\partial L}
\nonumber \\ &=
-\frac{\alpha}{\pi (\alpha+2) L}\;\frac{\partial T_r}{\partial L}\,,
\quad T_r=\frac{2 \pi}{\omega_r}\;.
\label{curvtr}
\end{align}
According to Eqs.~(\ref{freq}) and (\ref{fldef}) with the help of 
Ref.~\cite{byrdfried}, $\omega_r^{-1}$ is obviously simpler quantity 
to differentiate over $L$ than $f(L)$,
\begin{equation}
\omega_r^{-1}=
\frac{1}{2 \pi \sqrt{2}}\,\int_{x_{\rm min}}^{x_{\rm max}} 
\frac{\rmd x}{\sqrt{Q(x,L,\alpha)}},
\label{barom}
\end{equation}
\begin{equation}
Q(x,L,\alpha)=\left(1-x^{\alpha/2}\right)\,x -L^2/2\;,
\label{Qx}
\end{equation}
and $x=r^2$.  The turning points
$x_{\rm min}(L,\alpha)$ and 
$x_{\rm max}(L,\alpha)$ are determined by the equation:
\begin{equation}
Q(x,L,\alpha)=0\;.
\label{turneq1}
\end{equation}
Thus, we may calculate $\omega_r^{-1}$
and $f(L)$, and then, use Eqs.~(\ref{actra}) and (\ref{curvtr}) 
for the radial action
$I_r$, and curvature $K$ at the scaled energy $\varepsilon=1$.
Then, one obtains their energy dependence through the scaling
equations (\ref{scalE}) and (\ref{scaliolk}), respectively. 

\section{Full analytical classical dynamics for powers 4 and 6}
\label{appC}

For the powers $\alpha=4$ and $6$,  the roots of function (\ref{Qx}), 
in particular, the turning points
$x_{\rm min}$ and $x_{\rm max}$ can be obtained explicitly
analytically.  Therefore, 
one can find the explicit analytical
expressions for the key quantities of the classical dynamics for the POT, 
namely,
the radial frequency $\omega_r$ [Eq.~(\ref{freq})] 
(or the radial period $T_r$), 
and the frequency ratio 
$f(L)$ [Eq.~(\ref{fldef})] in terms
of the elliptic integrals from 
Ref.~\cite{byrdfried} (all in dimensionless units).

For $\alpha=4$, one has the cubic polynomial equation 
$Q(x,L,\alpha)\equiv x- x^3 - L^2/2=0$ 
[Eqs.~(\ref{Qx}) and (\ref{turneq1})]
for the three roots $x_{\rm min}$, $x_{\rm max}$ 
and $x_1$; given by the Cardano formulas explicitly 
as functions of $L$ in the physical 
region $L \leq L_C$, 
$r_1 < 0 \leq r_{\rm min} \leq r_{\rm max}$; $x_q=r_q^2$. 
For the radial period $T_r$ [Eqs.~(\ref{curvtr}) and 
(\ref{freq})], 
one obtains the
analytical expression through these roots in terms of the complete 
elliptic integral $\mathrm{F}(\pi/2,\kappa)$ of the first kind 
\cite{trap,byrdfried},
\begin{equation}
T_r=\frac{2 \pi}{\omega_r}=
\frac{\sqrt{2}}{\sqrt{x_{\rm max}-x_1}}\;
\mathrm{F}\left(\frac{\pi}{2},\kappa\right),
\label{ominv4}
\end{equation}
where 
$\kappa=[\left(x_{\rm max}-x_{\rm min}\right)/\left(x_{\rm max}-x_1\right)]^{1/2}\;$.
For the ratio frequencies $f(L)$ [Eq.~(\ref{fldef})], one finds
\begin{equation}
f(L)=\frac{ L}{\pi \sqrt{2}\; x_{\rm max} \sqrt{x_{\rm max}-x_1}}\;
\Pi\left(\frac{r_{\rm max}-r_{\rm min}}{r_{\rm max}},\kappa\right)\;,
\label{fl4}
\end{equation}
where $\Pi(n,\kappa)$ is the complete elliptic integral of the 3rd kind 
\cite{byrdfried}.

For $\alpha=6$, one has the polynomial 
equation of the 4th power, 
$Q(x,L,6)\equiv x - x^4 -L^2/2=0$, having the 4 roots 
[two complex conjugated
$x_1 + i x_2$ and $x_1 - i x_2$, and again, two real positive 
roots, $x_{\rm min}$ and $x_{\rm max}$; see
Eqs.~(\ref{Qx}) and (\ref{turneq1})].
The radial period $T_r$ is determined through these 
roots by the expression
[similar to Eq.~(\ref{ominv4}], see Refs.~\cite{trap,byrdfried},
\begin{equation}
T_r=\frac{\sqrt{2}}{
\sqrt{AB\left(x_{\rm max}-x_1\right)}}\;
\mathrm{F}\left(\frac{\pi}{2},\kappa\right),
\label{ominv6}
\end{equation}
where 
$~\kappa=\{
[(x_{\rm max}-x_{\rm min})^2-(A-B)^2]/(4 \;AB)\}^{1/2}~$,
$~A=[\left(x_{\rm max}-x_1\right)^2+x_2^2]^{1/2}~$,
$~B=[(x_{\rm min}-x_1)^2+x_2^2]^{1/2}\,$.
(We reduced the 4-power polynomial equation to a cubic one
and obtained  its 4 analytically given roots, mentioned above, in the explicit 
Cardano's form as functions of $L$). For $f(L)$ [Eq.~(\ref{fldef})] 
at $\alpha=6$, one obtains \cite{byrdfried}
\begin{align}
& f(L)=\frac{\sqrt{2}\;L\;(A+B)}{\pi\;\sqrt{AB}
\left(Ax_{\rm min}-Bx_{\rm max}\right)}\; \nonumber \\ & \quad\times
\left[\beta\; \mathrm{F}\left(\frac{\pi}{2}\kappa\right) + 
\frac{\beta-\beta_1}{2(1-\beta^2)}\;
\Pi\left(\pi,\frac{\beta^2}{1-\beta^2},\kappa\right)\right]\;,
\label{tr6}
\end{align}
where $\Pi(\varphi,n,\kappa)$ is incomplete elliptic integral of the 3rd 
kind,
$~\beta=(A x_{\rm min} - B x_{\rm max})/(A x_{\rm min} + B x_{\rm max})~$, 
$\beta_1=(A-B)/(A+B)\;$. 
The curvatures $K$ [Eq.~(\ref{curvtr})] for $\alpha=4$ and $6$ are 
determined by taking analytically the derivative
of the radial period $T_r$ [Eqs.~(\ref{ominv4}) and (\ref{ominv6})] 
over $L$ through 
the derivatives of the roots $x_{\rm min}(L)$, $x_{\rm max}(L)$, 
$x_1(L)$ and $x_2(L)$ 
for the derivative
of $\mathrm{F}(\pi/2,\kappa)$ over $\kappa$ \cite{byrdfried}.
The expressions for the curvatures  $K$ at the both powers
$\alpha=4$ and $6$ can be found in the closed analytical
form through a rather bulky formulas, which contain the complete 
elliptic integrals of the 1st and 2nd kind.

\section{Classical dynamics and boundaries for the diameters}
\label{appD}

For the primitive diameter $D=(2,1)$, the action $S_D$ 
(all in this appendix in dimensionless units)
is specified analytically through the scaled period 
$\tau_{{}_{\!D}}$ and energy $\varepsilon$ by
\begin{equation}
S_D=\tau_D \varepsilon\;, \quad 
\tau_{{}_{\!D}}=\frac{4 \sqrt{2\pi}}{\alpha+2}\;
\Gamma\left(\frac{1}{\alpha}\right)\;\Gamma\left(
\frac{1}{2}+\frac{1}{\alpha}\right)\;,
\label{tauraD}
\end{equation}
where $\Gamma(x)$ is the Gamma function of a real positive 
argument $x$.
For the diameter PO boundaries, one can use the
same $L_{-}=0$, but $L_{+}=b^{}_{D}L_C$,  where 
\begin{equation}
b^{}_{D}= 1-\frac12\;\exp\left[-\left(\frac{L^{HO}_D-L_D}{2 \varDelta_D}
\right)^2\right] 
\label{bcoefD}
\end{equation}
(see Ref.~\cite{maf} 
and more details in relation to the HO limit in
Sec.~\ref{sec:holimit}),
$L_D=0\;$ is the stationary point,  
$L_D^{HO}=L_C/2=\varepsilon/(2\sqrt{2})\;$
is the upper angular momentum $L_{+}$ for the $D$ orbits
in the limit $\alpha \to 2$, in which 
$b^{}_{D} \rightarrow 1/2$.  In the
semiclassical limit $\varepsilon \gg 1$, one has 
 $b_D \rightarrow 1$. 
$~\varDelta_D=(\pi M n_r K_D)^{-1/2}~$ is the 
Gaussian width of the 
transition region between these two asymptotic limits.
The $D$ curvature for $\alpha \geq 2$ at $L=L_D$ is given by
\begin{equation}
K_D=
\frac{\Gamma\left(1-1/\alpha\right)}{\varepsilon \sqrt{2 \pi}\,
\Gamma\left(1/2-1/\alpha\right)}\;.
\label{curvraD}
\end{equation}
This exact analytical expression for the curvature 
$K_D$  at any $\alpha$ was derived by
using a power expansion in Eqs.\ (\ref{turneq1}) and 
(\ref{actra}) over the variable proportional to  
$L^2$ near $L=0$ up to the terms linear in $L^2$. 
The Maslov phase for the diameter orbit
was determined by Eq.~(\ref{maslra1}) at $n_r=2$ and $n_\varphi=1$. 
Note that for the limit $\alpha \rightarrow 4$, the general expressions for the
period $\tau_{{}_{\!D}}$ and  action $S_D$ [Eq.~(\ref{tauraD})],
the Maslov index $\sigma_D$ [Eq.~(\ref{maslra1})] 
with the same asymptotic (SSPM) limit of the constant part of the phase 
$\phi^{(D)}_d=-\pi (4 b_D-3)/4 \rightarrow -\pi/4$,  
and the curvature $K_D$ [Eq.~(\ref{curvraD})] 
for the diameter $(2,1)$ are identical
to those obtained in Ref.~\cite{trap}.

\section{The boundaries and curvature for circle orbits}
\label{appE}

For the arguments $\mathcal{Z}_p^{(\pm)}$ and $\mathcal{Z}_r^{(\pm)}$ 
of the error functions in Eq.~(\ref{ampra0}), one originally has
\begin{gather}
\mathcal{Z}_{p,MC}^{(\pm)}=\sqrt{-\frac{i}{2\hbar}\,
\mathcal{J}_{MC}^{(p)}}\,\left(p_r^{(\pm)}-p_r^*\right),\nonumber\\
\mathcal{Z}_{r,MC}^{(\pm)}=\sqrt{-\frac{i}{2\hbar}\,
\mathcal{J}_{MC}^{(r)}}\,(r^{(\pm)}-r_{{}_{\!C}}),
\label{argerrorC}
\end{gather}
where $p_r^*=0$ is the stationary point, 
$p_r^{(\pm)}$ and  $r^{(\pm)}$ are maximal and minimal
classically accessible values of $p_r$ and $r$ as the finite
integration limits for the corresponding variables.
To express the integration boundaries (\ref{argerrorC}) in an invariant form
through the 
curvature $K_C$ (\ref{curvraC}), and stability factor 
$F_{MC}$ (\ref{fgutzC}), 
one may use now the simple standard Jacobian transformations,
and the definition of the  
angle variable $\varTheta_r^\prime$ as 
canonically conjugated one with respect to the radial action
variable $I_r$ by means of the corresponding generating function.
In these transformations, we apply simple linear relations:
$~p_r^{\prime\prime}-p_r^* =
\left(\partial p_r^{\prime\prime}/\partial L\right)^*\,
\left(L-L^*\right)~$, and 
$~r^\prime-r^*= 
 \left(\partial r^\prime/\partial \varTheta_r^\prime\right)^*\,
\left(\varTheta_r^\prime-\varTheta_r^*\right)$, 
where we immediately recognize the Jacobian coefficients.  Note that
there is no crossing terms due to the isolated stationary point
$I_r^*=0,\;\varTheta_r^*=0$ and to equations for the canonical
transformations.  At the stationary point for the isolated circle
PO, one has $f(L_C)=-(\partial I_r/\partial
L)_{L=L_C}=-1/\sqrt{\alpha+2}$ [Eqs.~(\ref{fldef}), (\ref{omtcra}) and
(\ref{omc})].  For the transformation of the derivative $\partial
r^{\prime\prime}/\partial \varTheta_r^\prime$, one can apply the Liouville
conservation of the phase space volume for the canonical variables to
arrive at
$\partial r^{\prime\prime}/\partial \varTheta_r^\prime=
(\partial I_r/\partial L)/(\partial p_r^{\prime}/\partial L)$ and
$\left|\mathcal{ J}_{\rm CT}(p^{\prime}_r,p^{\prime\prime}_r)\right|=
|(\partial p_r^{\prime\prime}/\partial L)/(\partial p_r^\prime/\partial L)|=1$ 
at the PO conditions 
$r' \to r^{\prime\prime} \to r_{{}_{\!C}},\, p_r^\prime \to p_r^{\prime\prime} \to 0$.
Using also the Jacobian identity,
\begin{equation}
F_{MC}=
-\mathcal{J}_{MC}^{(p)}\mathcal{J}_{MC}^{(r)}/\mathcal{J}_{MC}\left(
p_r^\prime,p_r^{\prime\prime}\right)\;, 
\label{stabfactjac}
\end{equation}
one obtains Eq.~(\ref{limra0}) for the arguments of the error functions 
in Eq.~(\ref{ampra0}).

The expression (\ref{curvraC}) 
for the $C$ curvature $K_C$ (in dimensionless units 
at $\varepsilon=1$)
was obtained
from expansion of $f(L)$ [Eq.~(\ref{fldef})] 
as function of $L$  
in powers of $L_C -L=\epsilon^2$ up to the 2nd order terms
in $\epsilon$.  For this purpose, by using standard perturbation 
theory, we have to solve first  
Eq.~(\ref{turneq1}) for the turning points $r_{\rm min}$ and $r_{\rm max}$,
[the integration
limits in Eq.~(\ref{fldef})]
in the following general form ($r$ is
taken below in units of $R_0$),
\begin{eqnarray}
r_{\rm max}&=&r_C + c_1 \epsilon + c_2 \epsilon^2 + c_3 \epsilon^3
+ c_4 \epsilon^4 + \cdots,
\nonumber\\
r_{\rm min}&=&r_C - c_1 \epsilon + c_2 \epsilon^2 -  c_3 \epsilon^3
+ c_4 \epsilon^4 + \cdots.
\label{turnexp}
\end{eqnarray}
Existence of such form of the solutions follows from a symmetry
of the equation (\ref{turneq1}) with respect to the change of the sign of
$\epsilon$.
Substituting
these solutions into Eq.~(\ref{turneq1}) for arbitrarily small $\epsilon$,
one gets the system of the recurrent
equations for the coefficients $c_n$. 
The solutions of this system up to the 4th order in a 
perturbation parameter $\epsilon$ is given by
\begin{gather}
c_1=\sqrt{\frac{L_C}{\alpha}}, \quad
c_2=-\frac{\alpha+1}{6}\,c_1^2, \quad
c_3=\frac{(\alpha-2)(2\alpha+5)}{72}\,c_1^3, \nonumber\\
c_4=-\frac{(\alpha+1)(4\alpha^2+8\alpha+13)}{1080}\,c_1^4\;,
\label{cnsol}
\end{gather}
and so on.  We transform now the integration variable $r$ in the
integral of Eq.~(\ref{fldef}) for $f(L)$ to $y$,
$r=r_{{}_{\!C}}(1-y)$, such that
\begin{equation}
f(L)=-\frac{L-\epsilon^2}{\pi}\,
\int_{y_{\rm min}}^{y_{\rm max}}\frac{\rmd y}{(1-y)\,\sqrt{Q(y,L,\alpha)}}\;.
\label{flint}
\end{equation}
Here, $Q(y,L,\alpha)$ is given by Eq.~(\ref{Qx}),
\begin{align}
&Q(y,L,\alpha) \nonumber \\ &\quad=
2\,r_C^2\,\left[1-\frac{2}{\alpha+2}\,(1-y)^\alpha -L_C^2 +
2L_C \epsilon^2 - \epsilon^4 \right] \nonumber\\ &\quad
\equiv (y_{\rm max}-y)(y-y_{\rm min})\mathcal{R}(y)\; ,
\label{Qy}
\end{align}
\begin{align}
y_{\rm max}&=\bar{c}_1\epsilon -\bar{c}_2\epsilon^2 + 
\bar{c}_3\epsilon^3 - \bar{c}_4\epsilon^4, \nonumber \\
y_{\rm min}&=-\bar{c}_1\epsilon - \bar{c}_2\epsilon^2 - 
\bar{c}_3\epsilon^3 - \bar{c}_4\epsilon^4\;,
\label{yturnexp}
\end{align}
where $\bar{c}_n=c_n/r_{{}_{\!C}}$.
We use the last representation in Eq.~(\ref{Qy}), introducing a
new function $\mathcal{R}(y)$ of the new variable 
$y$ to separate the singularities of
the integrand in Eq.~(\ref{flint}) due to the turning points. 
This integrand has to be integrated exactly
by using  a smooth function $\mathcal{R}(y)$ of $y$,
which can be expanded in $y$
at $y=0$ up to the second order,
\begin{equation}
\mathcal{R}(y)=\mathcal{R}(0) + \mathcal{R}'(0)y + 
\frac{1}{2}\mathcal{R}''(0)y^2+\cdots.
\label{rfunexp}
\end{equation}
In order to get analytically the final result, 
we note that $y$ in this expansion is of the order of
$\epsilon$, according to Eq.~(\ref{yturnexp}). 
Substituting then these expansions (\ref{yturnexp}) and (\ref{rfunexp})
into very right of Eq.~(\ref{Qy}), we expand their middle
in $y$ at $y=0$ up to the 4th order.  After the cancellation
of $\epsilon^2$ from both sides, and simple algebraic transformations,
one has
\begin{gather}
\mathcal{R}(0)=\frac{2L_C}{\overline{c}_1^2}\left[1 +
\epsilon^2\,\overline{c}_2\,
\left(1- k_2\right)\right] + \mathcal{O}(\epsilon^4),\quad
k_2=\frac{\overline{c}_2}{\overline{c}_1^2},
\nonumber\\
\frac{\mathcal{R}'(0)}{\mathcal{R}(0)}=2k_2 + 
\mathcal{O}(\epsilon^2),\quad
\frac{\mathcal{R}''(0)}{\mathcal{R}(0)}=2k_2(3k_2 + 1) + \mathcal{O}(\epsilon^2)\;.
\label{rfunder}
\end{gather}
For the calculation of the circle orbit curvature $K_C$, we
obviously need only quadratic terms in $\epsilon$ [linear in ($L_C
-L$)].  Therefore, one may neglect the $\epsilon^2$ corrections in the
second and third lines of Eq.~(\ref{rfunder}) because they are
multiplied by $y\sim \epsilon$ and $y^2\sim \epsilon^2$ in the expansion
(\ref{rfunexp}), respectively.  Substituting now expansions
(\ref{yturnexp}) and (\ref{rfunexp}) into the integral over y in
Eq.~(\ref{flint}), and taking $\mathcal{R}(0)$ off the integral, one
then expands to the second order all quantities of the integrand in $y
\sim \epsilon$, except for $(y_{\rm max}-y)(y-y_{\rm min})$ under the
square root (in the denominator) which can be integrated exactly.
Taking remaining integrals as 
$\int {\rm}d y y^n/\sqrt{(y_{\rm max}-y)(y-y_{\rm min})}$ 
from $y_{\rm min}$ to $y_{\rm max}$ [Eq.~(\ref{yturnexp})],
and then, expanding finally $f(L)$
[Eq.~(\ref{flint})] in $\epsilon$, we find that the linear terms
exactly disappear.  It must be the case because $f(L)$ is an even
function of $\epsilon$. Thus, the coefficient in front of $\epsilon^2$
with the expressions for $\overline{c}_n$ ($n=1,2,3$) from
Eq.~(\ref{cnsol}) is Eq.~(\ref{curvraC}) for the curvature $K_C$.
We can also use this perturbation method for calculations of the
next order curvatures, for instance, $\partial^3I_r/\partial L^3$,
which appears 
in expansion of the phase integral in the
exponent up to the third order terms near the stationary points within a
more precise (3rd-order) ISPM \cite{spheroid}.


\end{document}